\documentclass[11pt,a4paper]{article}
\usepackage[margin=1.0in]{geometry}
\usepackage{amssymb,amsmath,amsthm,verbatim,mathtools,needspace,enumitem,etoolbox,graphicx,physics,microtype,afterpage,xspace,tabularx,lmodern,multirow,boldline, makecell, booktabs,mathrsfs}
\usepackage{soul}
\usepackage{appendix}
\usepackage{tensor}
\usepackage{array}
\usepackage[normalem]{ulem}
\usepackage{calligra, calrsfs}
\usepackage[mathscr]{euscript}
\usepackage[normalem]{ulem}
\usepackage[dvipsnames]{xcolor}
\usepackage{xr-hyper}
\definecolor{linkcolor}{rgb}{0.0,0.3,0.5}
\usepackage[unicode, colorlinks=true, linkcolor=linkcolor, citecolor=linkcolor, filecolor=linkcolor, urlcolor=linkcolor, linktocpage, breaklinks]{hyperref}
\usepackage[all]{hypcap}
\usepackage[T1]{fontenc}
\usepackage{slashed}
\usepackage[utf8]{inputenc}
\hypersetup{colorlinks=true,citecolor=magenta,linkcolor=violet,urlcolor=magenta}
\usepackage{multirow}

\usepackage{wasysym}
\usepackage{cite}
\usepackage{cancel}

\allowdisplaybreaks

\numberwithin{equation}{section}

\input{epsf}
\newcommand{\beq}{\begin{equation}}
\newcommand{\eeq}{\end{equation}}
\newcommand{\bea}{\begin{eqnarray}}
\newcommand{\eea}{\end{eqnarray}}

\newcommand{\tho}{\text{\thorn}}

\begin{document}

\begin{titlepage}

    
    \bigskip

    \begin{center}
        %
         {\Large\textbf{Tidal perturbations of an extreme mass ratio inspiral around a Kerr black hole}}
         \\
         \bigskip
        
        \line(2,0){450}
        \bigskip
          {\textbf{
           Marta Cocco,$^{* \, \dagger \,}$\footnote{marta.cocco@nbi.ku.dk}
            Gianluca~Grignani,$^{* \,}$\footnote{gianluca.grignani@unipg.it}
            Troels~Harmark,$^{\dagger \, \ddagger \,}$\footnote{harmark@nbi.ku.dk}
            Marta~Orselli,$^{ * \, \dagger \,,}$\footnote{marta.orselli@unipg.it}
            David Pereñiguez,$^{\S \,}$\footnote{dpereni1@jhu.edu}
            and~Maarten~van~de~Meent$^{~ \dagger \, \P \,} $\footnote{maarten.meent@nbi.ku.dk}
            }}
            
        \bigskip
        \medskip
        
        \textit{{}$^*$ Dipartimento di Fisica e Geologia, Universit\`a di Perugia, I.N.F.N. Sezione di Perugia, \\ Via Pascoli, I-06123 Perugia, Italy }\\
        \medskip
        \textit{{}$^\dagger$ Center of Gravity,  Niels Bohr Institute, Copenhagen University,\\ Blegdamsvej 17, DK-2100 Copenhagen \O{}, Denmark}\\ 
        \medskip
         \textit{{}$^\ddagger$ Nordita, KTH Royal Institute of Technology and Stockholm University,\\
  Hannes Alfv\'ens v\"ag 12, SE-106 91 Stockholm, Sweden}\\ 
        \medskip
        \textit{{}$^\S$ William H. Miller III Department of Physics and Astronomy,
Johns Hopkins University, 3400 North Charles Street, Baltimore, Maryland, 21218, USA}
        \\
        \medskip
        \textit{{}$^\P$ Max Planck Institute for Gravitational Physics (Albert Einstein Institute), Am M\"uhlenberg 1, Potsdam 14476, Germany}
        \line(2,0){450}
        \vspace{10mm}

\begin{abstract}
We determine the metric of a Kerr black hole subject to external tidal fields
using metric reconstruction techniques. Working within the Newman--Penrose
formalism, we solve the Teukolsky master equation for static, quadrupolar modes
associated with a slowly varying tidal environment, and reconstruct the
corresponding metric perturbation in the outgoing radiation gauge. 
As an application, we derive the secular Hamiltonian governing the
motion of a test particle in the tidally deformed Kerr spacetime and investigate
long-term tidal effects relevant to extreme-mass-ratio inspirals. In particular, we compute tidal-induced shifts of the innermost stable circular orbit and the
light ring. 
We find that these tidal corrections are strongly spin dependent, with significantly larger effects for retrograde orbits around rapidly rotating black holes. Our results provide a fully analytic framework for
studying tidal interactions and secular dynamics in rotating black-hole spacetimes, with direct applications to gravitational-wave modeling and tests of
gravity in the strong-field regime.
\end{abstract}

\vspace{4cm}

\vfill

\end{center}

\setcounter{footnote}{0}

\end{titlepage}


\tableofcontents


\section{Introduction}
\label{sec:Introduction}

A central goal in black-hole perturbation theory is to obtain a tractable description of gravitational, electromagnetic, and neutrino perturbations of the Kerr spacetime. The pioneering work of Teukolsky~\cite{Teukolsky:1972my, Teukolsky:1973ha}, based on the Newman--Penrose formalism~\cite{Newman:1961qr}, established separable wave equations for the Weyl scalars $\psi_0$ and $\psi_4$, which encode the behavior of Kerr metric perturbations in vacuum. Together with earlier developments in Kerr geometry~\cite{Carter:1968rr, Chandrasekhar:1985kt}, these results laid the mathematical foundations for the study of black-hole stability, tidal interactions, and superradiant phenomena~\cite{Press:1972zz, Teukolsky:1974yv}.

Working with Weyl scalars, rather than the metric perturbation itself, provides direct access to physically relevant observables such as energy and angular momentum fluxes~\cite{Chrzanowski:1975wv}. This framework includes prescriptions for quantifying the absorption of mass and angular momentum by a black hole exposed to external radiation~\cite{Poisson:2004cw}. The Teukolsky equation also enables studies of quasinormal modes and their couplings~\cite{Leaver:1985ax, Green:2022htq}. More broadly, it has been widely applied to problems ranging from the stability of the Kerr spacetime to waveform modeling, including hybrid approaches that combine black-hole perturbation theory with post-Newtonian expansions~\cite{Damour:2009wj}, as well as investigations of horizon absorption and tidal torquing in Kerr spacetimes~\cite{Chatziioannou:2012gq}.

However, some applications require explicit knowledge of the metric perturbation rather than the Weyl scalars alone. 
This is the case for the computation of gravitational self-force effects, which depend on local properties of the spacetime geometry~\cite{vandeMeent:2015lxa}, for second-order self-force calculations, which require the metric perturbation at first order as an input~\cite{Pound:2012nt}, and more generally for extensions of black-hole perturbation theory beyond the linear regime~\cite{Loutrel:2020wbw, Ripley:2020xby}.

Metric reconstruction techniques based on Hertz or Debye potentials~\cite{ Kegeles:1979an,PhysRevLett.41.203} allow one to recover the spacetime perturbation from the Weyl scalars. These methods are essential for the calculation of Geroch--Hansen multipole moments~\cite{Geroch:1973am, LeTiec:2020bos}, the construction of initial data for binaries of rotating black holes, and the development of waveform models~\cite{Yunes:2005ve}.

One important application concerns the tidal deformation of black holes, which in the linear regime is described by vacuum perturbations of the Kerr metric. Previous studies address fundamental questions related to the relativistic theory of tidal Love numbers, which vanish for Schwarzschild black holes but exhibit a richer structure in the spinning case, including dissipative contributions associated with horizon absorption~\cite{Binnington:2009bb, Hartle:1973zz, Poisson:2004cw, Chatziioannou:2012gq, Chia:2020yla, LeTiec:2020bos, OSullivan:2014ywd}. The nonvanishing dissipative tidal response of Kerr black holes enters directly into the modeling of gravitational-wave signals from compact binaries: it contributes to the phase evolution and waveform amplitude, thereby affecting both parameter estimation and tests of the nature of compact objects~\cite{Tagoshi:1997jy, Alvi:2001mx, Yunes:2009ef, Yunes:2010zj, Flanagan:2007ix, Hinderer:2007mb, Baiotti:2010xh, Chatziioannou:2020pqz, Cardoso:2017cfl, Cardoso:2021qqu}. Beyond astrophysical applications, tidal perturbations also provide insight into the geometry and dynamics of black holes, for example through changes in the horizon area~\cite{Poisson:2009qj}, couplings between spin and tidal fields~\cite{Poisson:2014gka}, and modifications of the geodesic structure of tidally deformed spacetimes~\cite{Pani:2015hfa, Cardoso:2021qqu}.

In this paper, we apply metric reconstruction techniques to determine the spacetime metric of a tidally deformed rotating black hole, up to quadrupole order in the tidal deformation. Specifically, we consider a Kerr black hole of mass $M$ and angular momentum $J$, whose world line $\mathscr{L}$ lies in an external spacetime influenced by tidal fields. We assume that these tidal fields vary slowly in space and time.

We should point out that metric reconstruction techniques for constructing the tidally deformed metric of a Kerr black hole have been extensively developed in the literature. Existing studies include extensions beyond the quadrupole approximation, treatments of dynamical tidal fields, and analyses in different gauges and reconstruction schemes~\cite{LeTiec:2020bos, Yunes:2005ve, Berens:2024czo, Berens:2025kkm}. However, many available results are either implicit or not immediately suited for direct astrophysical applications. In this context, we provide explicit, ready-to-use expressions for the metric perturbation in a regular gauge, sourced by slowly varying quadrupolar tidal fields. Our presentation is intended to offer a compact and practical framework that can be directly applied to problems involving strong-field (SF) tidal interactions and orbital dynamics.

Specifically, we are interested in deriving the metric in the near-zone region, defined as the region at distances $r$ from the world line such that $r \ll \mathcal{R}$, where $\mathcal{R}$ denotes the tidal length scale of the external spacetime. This requirement imposes the key condition that the size of the black hole, characterized by its mass $M$, must be small compared to the tidal scale $\mathcal{R}$, i.e., $M \ll \mathcal{R}$. This hierarchy ensures that the tidal deformation can be treated perturbatively and that, on scales larger than $\mathcal{R}$, the external spacetime is not significantly affected by the presence of the black hole~\cite{Poisson:2009qj}.

We apply our result for the metric perturbation of a Kerr black hole to study the secular dynamics of a test particle in the tidally deformed spacetime. This application is particularly relevant for Extreme-Mass-Ratio Inspiral (EMRI) systems, which consist of a stellar-mass compact object inspiraling into a supermassive black hole with a mass ratio of order $10^{-5}$--$10^{-6}$ or smaller~\cite{Berry:2019wgg, Barack:2018yly}. EMRIs can emit gravitational-wave signals lasting up to $\mathcal{O}(10^5)$ orbital cycles in the millihertz LISA band~\cite{Barack:2018yly}, making them natural laboratories for probing secular phenomena and prime targets for space-based detectors such as LISA.

We use the secular Hamiltonian obtained from the reconstructed metric perturbation to investigate the properties of two characteristic orbits for massive and massless particles in a tidally deformed spacetime, namely the Innermost Stable Circular Orbit (ISCO) and the light ring (LR), respectively. In particular, we compute the tidal-field-induced shifts in the location and properties of these two orbits, and analyze how these shifts depend on the spin of the rotating black hole. We find that both the ISCO and the LR experience spin-dependent tidal shifts whose magnitude and sign depend on the relative orientation between the orbital angular momentum and the black-hole spin. In particular, for a rapidly rotating black hole, retrograde orbits exhibit substantially larger tidal corrections than prograde ones.\footnote{We refer to orbits with angular momentum aligned (anti-aligned) with the black-hole spin as prograde (retrograde). In the literature, these are sometimes called co-rotating and counter-rotating orbits.}

The paper is organized as follows. In Sec.~\ref{Sec:Teukolsky}, we introduce the formalism for describing tidal perturbations of a Kerr black hole and present the Teukolsky master equation in the Hartle--Hawking tetrad, providing a general solution for static modes, which we then specialize to the case of slowly varying tidal fields. In Sec.~\ref{sec:MetricReconstruction}, we summarize the metric reconstruction procedure, and in Sec.~\ref{sec:explicit_Kerr_perturbations}, we apply it to the Kerr spacetime, providing explicit analytic expressions for the components of the metric perturbation at quadrupole order. In Sec.~\ref{sec:Secular_Dynamics}, we describe the procedure for computing the secular dynamics of a test particle in a tidally deformed spacetime and derive the secular Hamiltonian perturbatively to first order in the tidal fields, using the reconstructed metric obtained in the previous section. We then use these results in Sec.~\ref{sec:ISCO_LR_shifts} to study how the properties of the ISCO and the LR are affected by tidal fields. Finally, Sec.~\ref{Sec: Conclusions} summarizes our conclusions and outlines possible directions for future work.

Throughout this paper, we employ geometrized units with $G=c=1$ and adopt the ``mostly plus'' metric signature $(-,+,+,+)$. 
Greek indices run over spacetime coordinates, while Latin indices run over spatial coordinates.


\section{Teukolsky master equation}
\label{Sec:Teukolsky}
To compute perturbations of the Kerr metric induced by external tidal fields, we employ the {\it Teukolsky Master Equation} (TME), first introduced by Teukolsky in his pioneering works~\cite{Teukolsky:1973ha, Teukolsky:1974yv}. In this section, we briefly review the main ingredients of the Teukolsky formalism relevant to the construction of tidally deformed Kerr geometries.
The TME provides a powerful framework for analyzing linear perturbations of rotating black holes in terms of gauge-invariant curvature variables.

We start by considering a linear perturbation of the Kerr metric,
\begin{equation}
\label{eq:fullmetric}
g_{\mu\nu} = \tilde{g}_{\mu\nu} + h_{\mu\nu},
\end{equation}
where $\tilde{g}_{\mu\nu}$ denotes the metric of the unperturbed Kerr black hole, and $h_{\mu\nu}$ represents the tidal perturbation, which we compute explicitly up to quadrupole order.

We will work in ingoing Kerr coordinates $(v,r,z,\psi)$. These coordinates are related to the usual Boyer--Lindquist coordinates  $(t,r,\theta,\phi)$ through the transformation 
\begin{equation}
\label{eq:ingoing_coordinates}
v = t + r_*, 
\qquad 
\psi = \phi + r_\sharp, 
\qquad 
z = \cos\theta ,
\end{equation}
where  
$r_*$ and $r_\sharp$ are defined as
\begin{equation}
\label{eq:rstar}
r_* = r 
+ \frac{2 M r_+}{r_+ - r_-} 
\ln\!\left( \frac{r - r_+}{2M} \right)
- \frac{2 M r_-}{r_+ - r_-} 
\ln\!\left( \frac{r - r_-}{2M} \right) ,
\end{equation}
\begin{equation}
\label{eq:rsharp}
r_\sharp = \frac{a}{r_+ - r_-} 
\ln\!\left( \frac{r - r_+}{r - r_-} \right) .
\end{equation}
Here $r_\pm = M \pm \sqrt{M^2 - a^2}$ denote the outer and inner horizons of the Kerr spacetime, and $a = J/M$ is the spin parameter. In contrast to Boyer--Lindquist coordinates, ingoing Kerr coordinates remain regular at the event horizon. In these coordinates, the Kerr metric takes the form
\begin{align}
\label{eq:Kerr}
ds^2 ={}& - \left(1 - \frac{2 M r}{\Sigma} \right) dv^2 
+ 2 \, dv \, dr 
- 2 a (1 - z^2) \left( \frac{2 M r}{\Sigma} \, dv + dr \right) d\psi \notag \\
& + \frac{\Sigma}{1 - z^2} \, dz^2 
+ \frac{1 - z^2}{\Sigma} \left[ (r^2 + a^2)^2 - a^2 \Delta (1 - z^2) \right] d\psi^2 ,
\end{align}
where $\Sigma = r^2 + a^2 z^2 = \zeta \bar{\zeta}$, with $\zeta = r - i a z$, and $\Delta = r^2 - 2 M r + a^2$.

Teukolsky's key insight was to reformulate the linearized Einstein equations in terms of curvature perturbations rather than metric perturbations. Specifically, starting from the Weyl tensor $C_{\mu\nu\rho\sigma}$ and writing the metric as in Eq.~\eqref{eq:fullmetric}, one can extract the part of the Weyl tensor that depends linearly on the perturbation $h_{\mu\nu}$. Following Ref.~\cite{Berens:2024czo}, we denote this contribution by $C_{\mu\nu\rho\sigma}^{(1)}$. Using this quantity, one can define two complex scalars, $\psi_0$ and $\psi_4$, which encode the two independent degrees of freedom of the metric perturbation $h_{\mu\nu}$. They are defined as
\begin{subequations}
\label{eqs:WeylScalars}
\begin{align}
\label{eq:WeylScalars0}
\psi_0 & \equiv C_{\mu\nu\rho\sigma}^{(1)} \, l^\mu m^\nu l^\rho m^\sigma , \\
\label{eq:WeylScalars4}
\psi_4 & \equiv C_{\mu\nu\rho\sigma}^{(1)} \, n^\mu \bar{m}^\nu n^\rho \bar{m}^\sigma ,
\end{align}
\end{subequations}
where $(l^\mu, n^\mu, m^\mu, \bar{m}^\mu)$ is a Newman--Penrose null tetrad satisfying the normalization and orthogonality conditions
\begin{subequations}
\label{eqs:Tetrad}
\begin{gather}
l^\mu l_\mu = n^\mu n_\mu
= m^\mu m_\mu
= \bar{m}^\mu \bar{m}_\mu
= l^\mu m_\mu
= l^\mu \bar{m}_\mu
= n^\mu m_\mu
= n^\mu \bar{m}_\mu
= 0 , \\
-l^\mu n_\mu = m^\mu \bar{m}_\mu = 1 .
\end{gather}
\end{subequations}
A common choice for the Newman--Penrose tetrad is the Kinnersley tetrad 
$\{l^\mu_{\rm K}, n^\mu_{\rm K}, m^\mu_{\rm K}, \bar{m}^\mu_{\rm K}\}$~\cite{Kinnersley:1969zza}, 
which is aligned with the principal null directions of the Kerr spacetime. 
In ingoing Kerr coordinates $(v,r,z,\psi)$, it takes the form~\cite{Teukolsky:1973ha}
\begin{subequations}
\label{eq:KinnersleyTetrad}
\begin{align}
l^\mu_{\rm K} &=
\left(
2 \frac{r^2 + a^2}{\Delta},\,
1,\,
0,\,
\frac{2 a}{\Delta}
\right), \\
n^\mu_{\rm K} &=
\frac{1}{2 \Sigma}
\left(
0,\,
-\Delta,\,
0,\,
0
\right), \\
m^\mu_{\rm K} &=
\frac{\sqrt{1 - z^2}}{\sqrt{2}\,\bar{\zeta}}
\left(
i a,\,
0,\,
-1,\,
\frac{i}{1 - z^2}
\right), \\
\bar{m}^\mu_{\rm K} &=
\frac{\sqrt{1 - z^2}}{\sqrt{2}\,\zeta}
\left(
- i a,\,
0,\,
-1,\,
- \frac{i}{1 - z^2}
\right).
\end{align}
\end{subequations}
Although the Kinnersley tetrad is aligned with the principal null directions of the Kerr spacetime and has the convenient property that $l^\mu_{\rm K}$ is affinely parametrized, $\nabla_{l_{\rm K}} l_{\rm K} = 0$, it is singular at the event horizon. This singular behavior is a property of the tetrad itself and is independent of the coordinate system used to express it. As a consequence, the associated Weyl scalars $\psi_0$ and $\psi_4$ also diverge at the horizon, despite the regularity of the Weyl tensor. To avoid this artificial singularity and ensure regularity of the curvature variables at the future event horizon, we instead adopt the \emph{Hartle--Hawking} (HH) tetrad
$\{l^\mu_{\rm HH}, n^\mu_{\rm HH}, m^\mu_{\rm HH}, \bar{m}^\mu_{\rm HH}\}$~\cite{Hawking:1972hy}.

The null vectors $l^\mu_{\rm K}$ and $n^\mu_{\rm K}$ of the Kinnersley tetrad are related to those of the HH tetrad, $l^\mu_{\rm HH}$ and $n^\mu_{\rm HH}$, through the transformations
\begin{equation}
l^\mu_{\rm HH} = \frac{\Delta}{2(r^2+a^2)} \, l^\mu_{\rm K},
\qquad
n^\mu_{\rm HH} = \frac{2(r^2+a^2)}{\Delta} \, n^\mu_{\rm K},
\end{equation}
while the complex null vector $m^\mu$ is the same in both tetrads~\cite{Poisson:2004cw, LeTiec:2020bos}.
Explicitly, the HH tetrad takes the form
\begin{subequations}
\label{eq:NPHH}
\begin{align}
l^\mu_{\rm HH} &=
\left(
1,\,
\frac{\Delta}{2(r^2+a^2)},\,
0,\,
\frac{a}{r^2+a^2}
\right), \\
n^\mu_{\rm HH} &=
\left(
0,\,
-\frac{r^2+a^2}{\Sigma},\,
0,\,
0
\right), \\
m^\mu_{\rm HH} &=
\frac{\sqrt{1-z^2}}{\sqrt{2}\,\overline{\zeta}}
\left(
i a,\,
0,\,
-1,\,
\frac{i}{1-z^2}
\right), \\
\bar{m}^\mu_{\rm HH} &=
\frac{\sqrt{1-z^2}}{\sqrt{2}\,\zeta}
\left(
- i a,\,
0,\,
-1,\,
- \frac{i}{1-z^2}
\right).
\end{align}
\end{subequations}
Moreover, the Weyl scalars in the HH tetrad are related to those in the Kinnersley tetrad by~\cite{Chandrasekhar:1985kt}
\begin{equation}
\label{eq:weylscalars_HH}
\psi_0^{\rm HH} = \frac{\Delta^2}{4 (r^2 + a^2)^2} \, \psi_0^{\rm K},
\qquad
\psi_4^{\rm HH} = \frac{4 (r^2 + a^2)^2}{\Delta^2} \, \psi_4^{\rm K} .
\end{equation}

Since we work exclusively with the HH tetrad in what follows, we will omit the superscript ``HH'' from the Weyl scalars unless otherwise stated.

In ingoing Kerr coordinates $(v,r,z,\psi)$ and in the HH tetrad, the TME reads
\begin{equation}
\label{eq:TME}
\begin{split}
\Bigg(
& - a^2 (1 - z^2) \, \partial_v^2
- 2 (a^2 + r^2) \, \partial_v \partial_r
- 2 a \, \partial_v \partial_\psi
- 2 a \, \partial_r \partial_\psi
- \frac{1}{1 - z^2} \, \partial_\psi^2
- \Delta \, \partial_r^2 \\
& + 2 (i a s z - r) \, \partial_v
+ 2 \left[ (s - 1)(r - M) - \frac{2 s r \Delta}{a^2 + r^2} \right] \partial_r
- \partial_z \!\left[ (1 - z^2) \, \partial_z \right] \\
& - 2 s \left( \frac{2 a r}{a^2 + r^2} + \frac{i z}{1 - z^2} \right) \partial_\psi
+ \left( \frac{s^2 z^2}{1 - z^2} - s
+ 8 M r \, \frac{(s - 2)a^2 + 2 r^2}{(a^2 + r^2)^2} \right)
\Bigg)
\Psi^{(s)} = 4 \pi \Sigma \, T^{(s)} .
\end{split}
\end{equation}
Here $\Psi^{(s)}$ denotes the perturbation field of spin weight $s$, and $T^{(s)}$ is the associated source term, which vanishes in vacuum. This form is obtained by projecting the Teukolsky operator given in Eq.~(3.28) of Ref.~\cite{Cano:2023tmv} onto the HH tetrad. In particular, Eq.~\eqref{eq:TME} applies to gravitational perturbations, corresponding to spin weights $s=\pm 2$.
For $s=\pm2$, the Weyl scalars $\psi_0$ and $\zeta^4\psi_4$ provide particular solutions of the TME with spin weights $+2$ and $-2$, respectively. 
We note that the expression of the source term $T^{(s)}$ is also tetrad dependent and therefore its explicit expression differs from that given in Ref.~\cite{Teukolsky:1972my}. 
However, in this work, we focus on vacuum perturbations and set the source term to zero.

A remarkable feature of Eq.~\eqref{eq:TME} is that it is separable. Each spin-weighted field $\Psi^{(s)}$ can be expanded as
\begin{equation}
\label{eq:ModeDecomposition}
\Psi^{(s)}(v,r,z,\psi)
= \int d\omega \sum_{\ell = |s|}^{\infty} \sum_{m=-\ell}^{\ell}
e^{-i \omega v + i m \psi} \,
c_{\ell m}^{(s)}(\omega)\,
R_{\omega \ell m}^{(s)}(r)\,
S_{\omega \ell m}^{(s)}(z) \, ,
\end{equation}
where the radial mode functions $R_{\omega \ell m}^{(s)}(r)$ and the angular mode functions
$S_{\omega \ell m}^{(s)}(z)$ satisfy separate second-order ordinary differential equations,
\begin{subequations}
\begin{align}
\label{eq:RadialODE}
\Bigg[
& \Delta \frac{d^2}{dr^2}
- 2 \left( (s-1)(r-M) - \frac{2 s r \Delta}{a^2 + r^2} + i K \right) \frac{d}{dr}
- \mathcal{V}_{\rm rad}(r)
- \lambda^{(s)}_{\omega \ell m}
\Bigg]
R_{\omega \ell m}^{(s)}(r)
= 0 , \\
\label{eq:AngularODE}
\Bigg[
& \frac{d}{dz} \left( (1 - z^2) \frac{d}{dz} \right)
+ a^2 \omega^2 z^2
- \frac{(m + s z)^2}{1 - z^2}
- 2 a \omega s z
+ s + A
\Bigg]
S_{\omega \ell m}^{(s)}(z)
= 0 .
\end{align}
\end{subequations}
Here $\lambda^{(s)}_{\omega \ell m}$ is the separation constant, which must in general be determined numerically, and we define
\begin{equation}
\label{eq:Potentials}
\begin{split}
K &\equiv (r^2 + a^2)\omega - a m , \qquad
A \equiv \lambda^{(s)}_{\omega \ell m} + 2 a m \omega - (a \omega)^2 , \\
K_s &\equiv K - (2 s - 1) a m , \qquad
\mathcal{V}_{\rm rad}(r) \equiv
\frac{2 i r K_s}{a^2 + r^2}
+ 8 M r \frac{(s-2)a^2 + 2 r^2}{(a^2 + r^2)^2} .
\end{split}
\end{equation}
For $\omega \neq 0$, the angular equation defines the spin-weighted spheroidal harmonics.


\subsection{Static modes}
We restrict our analysis to \emph{slowly varying tidal fields}, so that the TME~\eqref{eq:TME} can be solved for static modes, i.e., with $\omega = 0$. In this case, Eq.~\eqref{eq:AngularODE} for the angular mode functions is satisfied by the spin-weighted spherical harmonics ${}_s Y_{\ell m}(z,\psi)$, and the separation constant reduces to~\cite{vandeMeent:2015lxa, Berens:2024czo}
\begin{equation}
\lambda_{\ell m}^{(s)} \equiv \lambda_{0 \ell m}^{(s)}
= \ell(\ell+1) - s(s+1) .
\end{equation}
The spin-raising operator $\eth_s$ and the spin-lowering operator $\bar{\eth}_s$ can be written as
\begin{align}
\label{eq:spin_s_raising_op}
\eth_s &\equiv \frac{1}{\sqrt{1 - z^2}}
\left( (1 - z^2)\,\partial_z - i\,\partial_\psi + s z \right), \\
\bar{\eth}_s &\equiv \frac{1}{\sqrt{1 - z^2}}
\left( (1 - z^2)\,\partial_z + i\,\partial_\psi - s z \right).
\end{align}
They allow to construct the spin-weighted spherical harmonics ${}_s Y_{\ell m}$ recursively from the scalar spherical harmonics
$Y_{\ell m} \equiv {}_0 Y_{\ell m}$, for $\ell \ge |s|$, according to
\begin{alignat}{3}
\label{eq:SWSphericalHarmonics}
{}_{s+1}Y_{\ell m}
&\equiv \left[(\ell - s)(\ell + s + 1)\right]^{-1/2}
\, \eth_s({}_s Y_{\ell m}),
&&\qquad \text{for $s \ge 0$}, \\
{}_{s-1}Y_{\ell m}
&\equiv -\left[(\ell + s)(\ell - s + 1)\right]^{-1/2}
\, \bar{\eth}_s({}_s Y_{\ell m}),
&&\qquad \text{for $s \le 0$}.
\end{alignat}
Here
\begin{equation}
\label{eq:SphericalHarmonics}
Y_{\ell m}(z,\psi)
\equiv
\sqrt{\frac{(2\ell + 1)(\ell - m)!}{4\pi (\ell + m)!}}\,
P_\ell^m(z)\, e^{i m \psi}
= (-1)^m \, \bar{Y}_{\ell,-m}(z,\psi),
\end{equation}
where $P_\ell^m(z)$ are the associated Legendre polynomials of degree $\ell$ and order $m$.
The spin-weighted spherical harmonics ${}_s Y_{\ell m}(z,\psi)$ form an orthonormal basis on the unit two-sphere $\mathbb{S}^2$, satisfying the orthogonality relation
\begin{equation}
\oint_{\mathbb{S}^2} {}_s Y_{\ell m} \, {}_s \bar{Y}_{\ell' m'} \, d\Omega
= \delta_{\ell \ell'} \, \delta_{m m'} .
\end{equation}
The radial equation~\eqref{eq:RadialODE} for static modes,
$R^{(s)}_{\ell m} \equiv R^{(s)}_{0 \ell m}$, then reduces to
\begin{align}
\label{eq:radialODEstatic}
& x (x+1) \, \frac{d^2 R^{(s)}_{\ell m}}{d x^2}
+ \left[
2 i \gamma m
- (2x+1)(s-1)
+ \frac{4 s x (x+1)(\delta + x)}{\gamma^2 + (\delta + x)^2}
\right]
\frac{d R^{(s)}_{\ell m}}{d x}
\notag \\
& \quad
+ \left[
\frac{
4 (\delta + x) \bigl(i \gamma m s - 4 \delta + 2\bigr)
\bigl(\gamma^2 + (\delta + x)^2\bigr)
- \gamma^2 (2 \delta - 1)(s-4)
}{
\bigl(\gamma^2 + (\delta + x)^2\bigr)^2
}
- \lambda^{(s)}_{\ell m}
\right]
R^{(s)}_{\ell m}(x)
= 0 \, .
\end{align}
Here, we have introduced the dimensionless variables
\begin{equation}
\label{eq:x_and_gamma_and_delta}
x \equiv \frac{r - r_+}{r_+ - r_-},
\qquad
\gamma \equiv \frac{a}{r_+ - r_-},
\qquad
\delta \equiv \frac{r_+}{r_+ - r_-} .
\end{equation}
As already mentioned, we restrict our analysis to the gravitational case $s=\pm 2$.


\section{Metric reconstruction}
\label{sec:MetricReconstruction}

As mentioned in the Introduction, metric reconstruction techniques have been widely used to derive metric perturbations of spinning black holes~\cite{LeTiec:2020bos, Yunes:2005ve, Berens:2024czo, Berens:2025kkm}. In this section, we summarize the metric reconstruction procedure relevant for our analysis and specialize it to slowly varying quadrupolar tidal fields.
The reconstruction of the metric perturbation from solutions of the TME relies on the introduction of two tensorial differential operators, ${\mathcal{S}_0^\dag}_{\mu\nu}$ and ${\mathcal{S}_4^\dag}_{\mu\nu}$,
which, when applied to appropriate scalar potentials, yield the corresponding
metric perturbation in a radiation gauge. These scalar quantities, known as \emph{Hertz potentials}, satisfy an adjoint version of the TME and serve as intermediaries between the Weyl scalars $\psi_0$, $\psi_4$,
and the metric perturbation $h_{\mu\nu}$
\cite{Chrzanowski:1975wv, Wald:1973wwa, Berens:2024czo, LeTiec:2020bos}. \footnote{We remind the reader that we work in the HH tetrad and that, as stated below Eq.~\eqref{eq:weylscalars_HH}, we omit the superscript ``HH'' on the Weyl scalars $\psi_0$ and $\psi_4$ throughout the remainder of the paper.}

For a solution $\Psi_{\rm H}$ of the adjoint Teukolsky equation associated with
the spin-$s=-2$ field, the metric perturbation in the \emph{ingoing radiation gauge} (IRG) is given by
\begin{equation}
\label{eq:IRG}
h_{\mu\nu}^{\rm IRG}
= 2\,\mathrm{Re}\!\left( {\mathcal{S}_0^\dag}_{\mu\nu}\,\Psi_{\rm H} \right)
= 2\,\mathrm{Re}\!\left( {\mathcal{S}_0^\dag}_{\mu\nu}\,\Psi^{(-2)} \right),
\end{equation}
which satisfies the gauge conditions $l^\mu h_{\mu\nu}^{\rm IRG}=0$ and
$g^{\mu\nu} h_{\mu\nu}^{\rm IRG}=0$.

Similarly, for a Hertz potential $\Psi'_{\rm H}$ associated with the spin-$s=+2$
field, the metric perturbation in the \emph{outgoing radiation gauge} (ORG)
reads
\begin{equation}
\label{eq:ORG}
h_{\mu\nu}^{\rm ORG}
= 2\,\mathrm{Re}\!\left( {\mathcal{S}_4^\dag}_{\mu\nu}\,\Psi'_{\rm H} \right)
= 2\,\mathrm{Re}\!\left( {\mathcal{S}_4^\dag}_{\mu\nu}\,\zeta^4 \Psi^{(+2)} \right),
\end{equation}
and satisfies $n^\mu h_{\mu\nu}^{\rm ORG}=0$ and
$g^{\mu\nu} h_{\mu\nu}^{\rm ORG}=0$.

The Hertz potentials $\Psi_{\rm H}$ and $\Psi'_{\rm H}$ are related to the Weyl
scalars through fourth-order differential operators. In particular, the fields $\Psi^{(+2)}$ and $\Psi^{(-2)}$ appearing in Eqs.~\eqref{eq:ORG} and~\eqref{eq:IRG} are auxiliary Teukolsky fields of spin weight $+2$ and $-2$, respectively, and should not be confused with the Weyl scalars $\psi_0$ or $\zeta^4\psi_4$ \cite{vandeMeent:2015lxa, LeTiec:2020bos, Berens:2024czo}.

The explicit form of the reconstruction operators
${\mathcal{S}_0^\dagger}_{\mu\nu}$ and ${\mathcal{S}_4^\dagger}_{\mu\nu}$
can be written compactly within the Geroch--Held--Penrose (GHP) formalism~\cite{Geroch:1973am},
which refines the Newman--Penrose approach~\cite{Newman:1961qr}
by introducing derivative operators with well-defined spin and boost weights,
denoted by $\tho$, $\tho'$, $\eth$, and $\eth'$.

In the Newman--Penrose formalism, the connection associated with the null tetrad
is encoded in twelve complex spin coefficients
$(\kappa, \rho, \sigma, \tau, \kappa', \rho', \sigma', \tau', \epsilon, \beta, \epsilon', \beta')$.
In the GHP formalism, these coefficients enter naturally through the derivative operators, which act on quantities of GHP weight $(p,q)$ according to\footnote{We recall that a quantity $\eta$ has weights $(p,q)$ if under a frame transformation $(l,n,m)\mapsto(\lambda \bar{\lambda}l,\lambda^{-1} \bar{\lambda}^{-1}n,\lambda \bar{\lambda}^{-1}m)$ with $\lambda\in\mathbb{C}-\{0\}$ then $\eta$ transforms as $\eta\mapsto \lambda^{p}\bar{\lambda}^{q}\eta$.} 
\begin{equation}
\label{eq:GHP_derivatives}
\tho = \mathcal{D} - p \epsilon - q \overline{\epsilon} , \qquad
\eth = \delta - p \beta + q \overline{\beta}' , \qquad
\tho' = \mathcal{D}' + p \overline{\epsilon} + q \overline{\epsilon}' , \qquad
\eth' = \delta' + p \beta' - q \overline{\beta} ,
\end{equation}
where the frame derivatives are defined as
\begin{equation}
\mathcal{D} = l^\mu \nabla_\mu , \qquad
\mathcal{D}' = n^\mu \nabla_\mu , \qquad
\delta = m^\mu \nabla_\mu , \qquad
\delta' = \bar{m}^\mu \nabla_\mu .
\end{equation}
In this notation, the reconstruction operators
${\mathcal{S}_0^\dag}_{\mu\nu}$ and ${\mathcal{S}_4^\dag}_{\mu\nu}$
used in Eqs.~\eqref{eq:IRG} and~\eqref{eq:ORG} read \cite{Kegeles:1979an,PhysRevLett.41.203} (see also \cite{Berens:2024czo}) 
\begin{subequations}
\begin{align}
\label{eq:GHP0}
{\mathcal{S}_0^\dag}_{\mu\nu}\Psi
&=
-\frac{1}{2}\,l_\mu l_\nu
\bigl(\eth-\tau\bigr)\bigl(\eth+3\tau\bigr)\Psi
-\frac{1}{2}\,m_\mu m_\nu
\bigl(\tho-\rho\bigr)\bigl(\tho+3\rho\bigr)\Psi
\notag\\
&\quad
+\frac{1}{2}\,l_{(\mu} m_{\nu)}
\Bigl[
\bigl(\tho-\rho+\overline{\rho}\bigr)\bigl(\eth+3\tau\bigr)
+\bigl(\eth-\tau+\overline{\tau}'\bigr)\bigl(\tho+3\rho\bigr)
\Bigr]\Psi , \\
\label{eq:GHP4}
{\mathcal{S}_4^\dag}_{\mu\nu}\Psi
&=
-\frac{1}{2}\,n_\mu n_\nu
\bigl(\eth'-\tau'\bigr)\bigl(\eth'+3\tau'\bigr)\Psi
-\frac{1}{2}\,\bar{m}_\mu \bar{m}_\nu
\bigl(\tho'-\rho'\bigr)\bigl(\tho'+3\rho'\bigr)\Psi
\notag\\
&\quad
+\frac{1}{2}\,n_{(\mu} \bar{m}_{\nu)}
\Bigl[
\bigl(\tho'-\rho'+\overline{\rho}'\bigr)\bigl(\eth'+3\tau'\bigr)
+\bigl(\eth'-\tau'+\overline{\tau}\bigr)\bigl(\tho'+3\rho'\bigr)
\Bigr]\Psi .
\end{align}
\end{subequations}
In the HH tetrad~\eqref{eq:NPHH}, the relevant spin coefficients are
\begin{subequations}
\label{eq:spincoefficients}
\begin{align}
\tau &=
-\frac{i a \sqrt{1-z^2}}{\sqrt{2}\,\Sigma},
\qquad
\tau'=
-\frac{i a \sqrt{1-z^2}}{\sqrt{2}\,\zeta^2},
\qquad
\rho=
\frac{\Delta}{2 (r^2+a^2)\,\zeta},
\qquad
\rho'=
\frac{r^2+a^2}{\zeta\,\Sigma},
\\
\beta &=
\frac{z}{2\sqrt{2(1-z^2)}\,\overline{\zeta}},
\qquad
\beta'=
\frac{r z + i a (z^2-2)}{2\sqrt{2(1-z^2)}\,\zeta^2},
\qquad
\epsilon=
\frac{M (r^2-a^2)}{2 (r^2+a^2)^2},
\qquad
\epsilon'=
\frac{a (a+i r z)}{\zeta\,\Sigma}.
\end{align}
\end{subequations}
The remaining spin coefficients $(\kappa,\kappa',\sigma,\sigma')$
vanish identically in the HH tetrad.
In the following, we work with perturbations in the ORG, Eq.~\eqref{eq:ORG}.
In this gauge, the metric perturbation naturally takes the form of a horizon-adapted
lightcone gauge, which is particularly well-suited for describing tidal perturbations
of black holes and for implementing metric reconstruction techniques based on the
Hertz potential (see, e.g.,
Refs.~\cite{Chrzanowski:1975wv, Poisson:2004cw, Poisson:2009qj, vandeMeent:2015lxa}).


\subsection{Hertz potential}
\label{subsec:HertzPotential}
In the ORG, the relevant solution of the TME corresponds to a Hertz potential $\Psi'_{\rm H}$ with spin weight $s=+2$.

In this work, we restrict our analysis to the lowest multipolar order of the external tidal
field, namely the quadrupole order corresponding to $\ell=2$.
The Hertz potential $\Psi'_{\rm H}$ can therefore be decomposed as
\begin{equation}
\label{eq:PsiH_l2_sum}
\Psi'_{\rm H}
= \sum_{m=-2}^{2} \Psi_{m}^{(+2)} .
\end{equation}
Each $m$-mode takes the general form
\begin{equation}
\label{eq:PsiH_l2_modes_gen}
\Psi_{m}^{(+2)}
= c_{2m}(\tilde v,a)\, \zeta^4 \, R_{2m}(r)\, {}_2 Y_{2m}(z,\psi) ,
\end{equation}
where $R_{2m}(r)$ are radial mode functions solving Eq.~\eqref{eq:RadialODE} with
$\omega=0$, $s=2$, and $\ell=2$, ${}_2 Y_{2m}(z,\psi)$ are spin-weighted spherical harmonics,
and $c_{2m}(\tilde v,a)$ are complex coefficients encoding the external tidal field.

Here $\tilde v$ denotes a slowly varying time variable, in the sense that the tidal field
is assumed to vary on timescales much longer than the characteristic dynamical timescale
of the black hole. As a consequence, time derivatives are negligible compared to spatial
derivatives~\cite{Binnington:2009bb, LeTiec:2020bos}. Accordingly, the Teukolsky equation
is solved in the static ($\omega=0$) sector, and the coefficients $c_{2m}$ depend on
$\tilde v$ only parametrically.

In principle, the coefficients $c_{2m}$ could also depend on the black-hole spin $a$.
However, previous analyses based on asymptotic matching of the Weyl scalar $\psi_0$ to the
external tidal curvature~\cite{Poisson:2004cw, Chatziioannou:2012gq} have shown that the
tidal multipole moments characterizing the external universe are independent of the spin
of the perturbed black hole. As a result, the coefficients $c_{2m}$ carry no dependence on $a$ and reduce simply to $c_{2m}=c_{2m}(\tilde v)$. This reflects the fact that the coefficients $c_{2m}$ characterize the external tidal environment rather than the intrinsic properties of the perturbed black hole. As a consequence, they can be determined directly from the Schwarzschild limit, as shown below.

In Eq.~\eqref{eq:PsiH_l2_modes_gen}, the radial mode function is given by 
\begin{equation}
\label{eq:R2m_solution}
R_{2m}(r)
=
\frac{\Delta^2}{4 (r^2+a^2)^2}
\, \frac{\Gamma(3 + 2 i m \gamma)}{4!}
\, \frac{\mathbf{F}(-4,1,-1+2 i m \gamma; -x)}{x^2 (1+x)^2} ,
\qquad |m|\leq 2 ,
\end{equation}
where $\mathbf{F}(a,b,c;z)$ denotes the regularized hypergeometric function, and
$x$ and $\gamma$ are defined in Eq.~\eqref{eq:x_and_gamma_and_delta}. 
It is worth noting that $R_{2m}(r)$ is an algebraic function of the Boyer--Lindquist radius $r$; indeed, notice that $\mathbf{F}(-4,1,-1+2 i m \gamma; -x)$ reduces to a polynomial. The Boyer--Lindquist radial coordinate is physically well-defined, since its level sets are totally-geodesic surfaces. This fact underlies the vanishing of black-hole tidal Love numbers in general relativity, and it extends to the general harmonic case $\ell\ne2$, where $R_{\ell m}(r)$ reads
\begin{equation}
\label{eq:Rlm_solution}
R_{\ell m}(r)
\sim
\frac{1}{(r^2+a^2)^2}\mathbf{F}(-2-\ell,-1+\ell,-1+2 i m \gamma; -x)\, ,
\qquad |m|\leq \ell ,
\end{equation}
where $\mathbf{F}(-2-\ell,-1+\ell,-1+2 i m \gamma; -x)$ is again a polynomial. The expression in Eq.~\eqref{eq:R2m_solution} coincides with the result obtained in the Kinnersley tetrad~\cite{LeTiec:2020bos},
up to the expected rescaling by the square of the factor $\Delta/[2(r^2+a^2)]$ relating the
Kinnersley and HH tetrads, in agreement with Eq.~\eqref{eq:weylscalars_HH}.

The spin-weighted spherical harmonics ${}_2 Y_{2m}(z,\psi)$ appearing in
Eq.~\eqref{eq:PsiH_l2_modes_gen} follow from Eq.~\eqref{eq:SWSphericalHarmonics}
with $s=2$ and $\ell=2$, and read
\begin{equation}
{}_2 Y_{2m}(z,\psi)
= \frac{1}{2\sqrt{6}} \, \eth_1 \eth_0 \, Y_{2m}(z,\psi),
\qquad |m|\leq 2 ,
\end{equation}
where the spin-raising operator $\eth_s$ is defined in
Eq.~\eqref{eq:spin_s_raising_op}, and $Y_{2m}(z,\psi)$ are the standard spherical
harmonics given in Eq.~\eqref{eq:SphericalHarmonics}.
Inserting this expression together with Eq.~\eqref{eq:R2m_solution} into
Eq.~\eqref{eq:PsiH_l2_modes_gen} yields
\begin{equation}
\label{eq:PsiH_l2_modes_def}
\Psi^{(+2)}_m
=
c_{2m}(\tilde v)\,
(r - i a z)^4\,
\frac{(r_+ - r_-)^4}{4 (r^2 + a^2)^2}\,
\frac{\Gamma(3 + 2 i m \gamma)}{4!}\,
\mathbf{F}(-4,1,-1+2 i m \gamma; -x)\,
{}_2 Y_{2m}(z,\psi) .
\end{equation}

The asymptotic behavior of the reconstructed metric must correspond to that of a tidally deformed black hole. This requires identifying the Weyl tensor in the intermediate region $r_+ \ll r \ll \mathcal{R}$ with the Weyl tensor of the external background spacetime responsible for the tidal deformation. Here $\mathcal{R}$ denotes the curvature scale of the external spacetime.

From the perspective of the external spacetime, the black hole can be treated as a point particle moving along a worldline $\mathscr{L}$, which is endowed with an orthonormal tetrad $(u^\alpha, e^\alpha_a)$. Here $u^\alpha$ is the black-hole four-velocity and $e^\alpha_a$ is a spatial triad parallel transported along $\mathscr{L}$. This tetrad defines the \emph{local asymptotic rest frame} of the black hole within the external geometry. Projecting the external Weyl tensor onto this frame yields the components~\cite{Zhang:1986cpa, Poisson:2009qj, Binnington:2009bb, Chatziioannou:2012gq, Poisson:2014gka}
\begin{equation}
\label{eq:Cij_and_Cijk}
C_{0b0d}
\equiv
C_{\alpha\beta\gamma\delta}
u^\alpha e^\beta_b u^\gamma e^\delta_d ,
\qquad
C_{0bcd}
\equiv
C_{\alpha\beta\gamma\delta}
u^\alpha e^\beta_b e^\gamma_c e^\delta_d .
\end{equation}
These frame components can in turn be recast in terms of two symmetric,
trace-free tensors $\mathcal{E}_{ab}$ and $\mathcal{B}_{ab}$, representing
the electric-type and magnetic-type quadrupole moments of the tidal field,
respectively~\cite{Poisson:2009qj}. They are defined as
\begin{equation}
\label{eq:Eab_and_Bab}
\mathcal{E}_{ab} \equiv C_{0a0b} , \qquad
\mathcal{B}_{ab} \equiv \frac{1}{2}\,\epsilon_{a}{}^{cd} \, C_{cdb0} ,
\end{equation}
where $\epsilon_{abc}$ denotes the three-dimensional Levi--Civita tensor.
Together, these tensors encode all ten independent components of the Weyl
tensor and therefore provide a complete description of the tidal environment
acting on the black hole.

In Refs.~\cite{Poisson:2009qj, Binnington:2009bb, Poisson:2004cw,
Chatziioannou:2012gq}, the coefficients $c_{2m}(\tilde v)$ were determined by
asymptotically matching the Weyl scalar $\psi_0$ to the tidal curvature in the
intermediate region $r_+ \ll r \ll \mathcal{R}$. This procedure was carried out
explicitly for a Schwarzschild black hole (for all multipolar orders $\ell$)
in Refs.~\cite{Poisson:2009qj, Binnington:2009bb}, and for a Kerr black hole up
to octupolar order in Refs.~\cite{Poisson:2004cw, Chatziioannou:2012gq}.

Taking advantage of the results of Refs.~\cite{Poisson:2009qj, Binnington:2009bb, Poisson:2004cw,
Chatziioannou:2012gq}, we can avoid the direct computation of the coefficients $c_{2m}$ through the matched asymptotic expansion procedure, and instead, we determine them by requiring that
the metric reconstructed from $\Psi'_{\rm H}$ reproduces the known quadrupolar
tidal perturbation computed in Ref.~\cite{Binnington:2009bb}, in the Schwarzschild limit $a=0$.~\footnote{This is done using the relation between the Weyl scalar $\psi_0$ and the
Hertz potential $\Psi'_{\rm H}$ (see, e.g.,
Eq.~(1.23a) of Ref.~\cite{Berens:2024czo}).} This prescription is justified by previous asymptotic-matching analyses of tidally perturbed Kerr black holes~\cite{Poisson:2004cw, Chatziioannou:2012gq}, which showed that the coefficients $c_{2m}$ are independent of the black-hole spin, as already discussed after Eq.~\eqref{eq:PsiH_l2_modes_gen}.


\section{Perturbation to the Kerr metric: results}
\label{sec:explicit_Kerr_perturbations}

In this section, we present the explicit form of the quadrupolar perturbation to the Kerr metric reconstructed in the ORG.
One of the main goals of this work is to provide compact and ready-to-use expressions for the reconstructed metric that can be directly used in astrophysical and SF applications. The perturbation is obtained by applying the reconstruction operator
${\mathcal{S}_4^\dag}_{\mu\nu}$, defined in Eq.~\eqref{eq:GHP4}, to the Hertz
potential with spin weight $s=+2$.
For each value of the azimuthal number $m$, we define the complex quantity
\begin{equation}
\label{eq:hmunu_complex_modes}
\hat{h}_{\mu\nu}^{(m)\,\rm ORG}
\equiv
{\mathcal{S}_4^\dag}_{\mu\nu}\,\Psi^{(+2)}_m ,
\end{equation}
where $\Psi^{(+2)}_m$ is defined in Eq.~\eqref{eq:PsiH_l2_modes_def}.

The full complex metric perturbation is then obtained by summing over the
allowed values of $m$,
\begin{equation}
\label{eq:hmunu_complex_full}
\hat{h}_{\mu\nu}^{\rm ORG}
=
\sum_{m=-2}^{2} \hat{h}_{\mu\nu}^{(m)\,\rm ORG} .
\end{equation}
The final step, as indicated in Eq.~\eqref{eq:ORG}, is to take the real part
of this expression, yielding
$h_{\mu\nu}^{\rm ORG} \equiv 2\,\mathrm{Re}\!\left[\hat{h}_{\mu\nu}^{\rm ORG}\right]$.
The explicit expressions for each $m$-mode are given below.
To simplify the notation, we henceforth omit the superscript ``ORG''.
We present the full expressions explicitly in order to facilitate direct reuse
in analytic and numerical applications. To this purpose, in the supplementary material, we also provide a \texttt{Mathematica} notebook with the components of the reconstructed metric. The notebook is also publicly available through the GitHub repository~\cite{KerrTidalGitHub}.

To present the $m=\pm1$ and $m=\pm2$ contributions in a compact form, we introduce the auxiliary symbols $\mathbf z$ and $\mathbf i$. These should not be interpreted as redefinitions of the coordinate $z=\cos\theta$ or of the imaginary unit $i$, but only as shorthand notation encoding sign dependence on $m$ through Eqs.~\eqref{eq:bolds_z_and_i_pm1} and~\eqref{eq:bolds_z_and_i_pm2}.

\subsection{\texorpdfstring{$m=0$}{m=0} mode}

We begin with the axisymmetric contribution $m=0$, which corresponds to the
purely polar part of the quadrupolar deformation. The corresponding components
$\hat{h}_{\mu\nu}^{(0)}$ are listed below.
\begin{subequations} 
    \label{eq:hmunu_m0}
    \begin{align}
        \hat{h}_{vv}^{(0)} & =  \frac{\mathcal{N}_{2 \, 0}}{\zeta \overline{\zeta}^3 } \Bigg[  a^6 z^4 (z^2-3) + 2 a^5 i z (z^2-1) \left(M (3 z^2+1) -3 r z^2 \right) \notag \\
        & + a^4 \left(2 M^2 z^2 (z^2-1)^2  - 2 M r (3 z^6 - 9 z^4 + 3 z^2 - 1)+3 r^2 z^4 (z^2-3)\right) \notag \\
        & - 2 a^3 i r z (z^2-1) \left(M^2 (6 z^2+2)-3 M r (3 z^2+1)+3 r^2 (z^2+1)\right) \notag \\
        & - a^2 r^2 \left(2 M^2 (3 z^4+1) - 4 M r (z^4+4 z^2-1)+3 r^2 (3 z^2 - 1)\right) \notag \\
        & -2 a i r^3 z (z^2-1) (4 M^2-8 M r+3 r^2) -r^4 (3 z^2-1) (2M - r)^2\Bigg] \,, \\
        \hat{h}_{vz}^{(0)} & =- \frac{ i~ \mathcal{N}_{2 \, 0}}{ \overline{\zeta}^2}  \Bigg[ a^5 z^2 (z^2-1)+2 i a^4 z \left(2 M z^2 -r (2 z^2-1)\right)+a^3 (z^2-1) \left(2 M^2 z^2- r (2 M-r) (3 z^2-1) \right) \notag \\
        & -4 i a^2 r z \left(2 M^2 z^2-M r (3 z^2-1)+r^2 z^2\right) - a r^2 (z^2-1) \left(2M^2 - r^2\right)+2 i r^4 z (2 M-r) \Bigg]\,,  \\
        \hat{h}_{v\psi}^{(0)} & = 2 \frac{\mathcal{N}_{2 \, 0}}{\overline{\zeta}^3} (1-z^2) \Bigg[ i a^6 z^3 - a^5 \left( z^4 (r-M)+z^2 (2 M-3 r)+M \right) \notag \\
        & +i a^4 z \left(M^2 (z^2-1)- M r (6 z^2 - 4) + r^2 (4 z^2-3)\right) \notag \\
        & - a^3 r \left(M^2 (2 z^4-5 z^2-1) - M r (3 z^4 - 12 z^2 + 1)+r^2 (z^4-6 z^2+1)\right) \notag \\
        & i a^2 r^2 z \left(5 M^2 (z^2-1)-8 M r (z^2-1)+r^2 (3 z^2-4)\right) \notag \\ 
        & + a \left(r^3 (M-r) (z^2 (5 M-3 r)-M+r)\right) + i r^5 z (2 M-r) \Bigg] \,, \\
        \hat{h}_{vr}^{(0)} ~ & = ~ \hat{h}_{rr}^{(0)} ~ = ~ \hat{h}_{rz}^{(0)} ~ = ~ \hat{h}_{r\psi}^{(0)} ~ = ~ 0 \,, \\
        \hat{h}_{zz}^{(0)} & = \mathcal{N}_{2\, 0} \Bigg[ a^4 z^2+2 i a^3 M z+a^2 \left(2 M^2 z^2-r (2 M-r) (3 z^2+1) \right)-2 i a r^2 (3 M-2 r) z \notag \\
        & +r^2 \left(2 M^2-r^2\right) \Bigg] \,, \\
        \hat{h}_{z\psi}^{(0)} & = - \frac{\mathcal{N}_{2 \, 0}}{\overline\zeta^2}(1-z^2)  \Bigg[i a^6 z^2 + 2 a^5 \left(M z^3-r z (z^2-1)\right) + i a^4 \left(2 M^2 z^2-2 M r (3 z^2-1)+r^2 (4 z^2-1)\right) \notag \\
        & -a^3 \left(4 M^2 r z^3-2 r^2 (3 M-r) z (z^2-1) \right) + i a^2 r^2 \left(2 M^2 (z^2-1)-2 M r (3 z^2-1)+3 r^2 z^2\right) \notag \\
        & + 2 a M r^3 z (2 M-r) - i r^4 (2 M^2 - r^2) \Bigg]\,, \\
        \hat{h}_{\psi\psi}^{(0)} & = -\frac{\mathcal{N}_{2 \,0}}{\overline\zeta^3} (1-z^2)^2 \Bigg[ i a^7 z^3 - a^6 \left(2 M (2 z^2+1)-3 r z^2\right) - i a^5 z \left(2 M^2+ r(2 M-r)(2 z^2-3) \right) \notag \\
        & +a^4 \left(2 M^2 r (4 z^2+1) - 12 M r^2 z^2 + r^3 (6 z^2-1)\right) + i a^3 \left(4 M r^2 (M-r) z (z^2-2)+r^4 z (z^2-6)\right) \notag \\
        & +a^2 r^3 \left(4 M^2 r^3 z^2 - 2 M r^4 (4 z^2-1) + r^5 (3 z^2-2)\right) + i a r^4 (6 M^2+2 M r - 3 r^2) z + r^5 (2 M^2-r^2) \Bigg] \,,
    \end{align}
\end{subequations}
where we defined the complex coefficient
\begin{equation} 
    \label{eq:N_20}
    \mathcal{N}_{2 \,0}\equiv\frac{c_{2 \,0}}{16}\sqrt{\frac{15}{2 \pi }} \,.
\end{equation}

\subsection{\texorpdfstring{$m=\pm1$}{m=±1} modes}

These modes are obtained analogously from Eq.~\eqref{eq:hmunu_complex_modes}
with $m=\pm1$, and their explicit expressions are reported below.
\begin{subequations} 
    \label{eq:hmunu_m_pm1}
    \begin{align}
        \hat{h}_{vv}^{(\pm 1)} & = -\frac{\mathcal{N}_{2 \, \pm 1}} {{\overline{\zeta}^3}} \sqrt{1-\mathbf{z}^2}\Bigg[ 9 \mathbf{i} a^5 \mathbf{z}^3+3 a^4 \left(M (3 \mathbf{z}^4-3\mathbf{z}^3-5\mathbf{z}^2+\mathbf{z}-2)-3 r \mathbf{z}^2(\mathbf{z}^2 - \mathbf{z} -3)\right) \notag \\
        &+3 \mathbf{i} a^3 \Big(2 M^2 (\mathbf{z}^2+\mathbf{z}^3-2\mathbf{z}^2-1) +3 r^2 \mathbf{z} (\mathbf{z}^3+3 \mathbf{z}^2-3 \mathbf{z}-3)\Big) \notag \\
        & +a^2 \Big(4 M^3 (\mathbf{z}^2+\mathbf{z}+1)+6 M^2 r (7\mathbf{z}^3+\mathbf{z}^2 -4\mathbf{z} -1) -18 M r^2 (4\mathbf{z}^3+3\mathbf{z}^2-3\mathbf{z}-1) \notag \\
        & +9 r^3 (3\mathbf{z}^3+3\mathbf{z}^2-3\mathbf{z}-1)\Big) \notag \\
        & -3 \mathbf{i} a r \Big(4 M^3 \mathbf{z}+4 M^2 r (5 \mathbf{z}^2+\mathbf{z}-1) -2 M r^2 (14\mathbf{z}^2 +5\mathbf{z} -4)+3 r^3 (3 \mathbf{z}^2+\mathbf{z}-1)\Big) \notag \\
        & -9 r^3 (2M-r)^2 \mathbf{z} \Bigg]\,, \\
        \hat{h}_{vz}^{(\pm 1)} & = \pm \frac{\mathcal{N}_{2 \, \pm 1}} {{\overline{\zeta}^2}} \sqrt{\frac{1-\mathbf{z}}{1+\mathbf{z}}}\Bigg[ 3 \mathbf{i} a^5 \mathbf{z}^2 (2 \mathbf{z}+1)+3 a^4 \mathbf{z} \left(3 M \mathbf{z} (\mathbf{z}^2-1)-r (3 \mathbf{z}^3-4\mathbf{z}-2)\right) \notag \\
        & +\mathbf{i} a^3 \left(2M^2 \mathbf{z}^2 (3 \mathbf{z}^2+4 \mathbf{z}-1) - 6 M r (3 \mathbf{z}^4+6 \mathbf{z}^3-2 \mathbf{z}-1)+3 r^2 (3 \mathbf{z}^4+6 \mathbf{z}^3-2 \mathbf{z}-1)\right) \notag \\
        & +a^2 r \left(2 M^2 (12 \mathbf{z}^3+8\mathbf{z}^2 -2 \mathbf{z} -3)-9 M r (4 \mathbf{z}^3+3 \mathbf{z}^2-1)+6 r^2 \mathbf{z}^2 (2 \mathbf{z}+1)\right) \notag \\
        &-\mathbf{i} a r \left(4 M^3 (2 \mathbf{z}+1)+2 M^2 r (3 \mathbf{z}^2+4 \mathbf{z}-1)-3 r^3 \mathbf{z}(\mathbf{z}+2)\right) \notag \\
        &-3 r^4 (2 M-r) (2 \mathbf{z}+1) \Bigg]\,,  \\
        \hat{h}_{v\psi}^{(\pm 1)} & = \frac{\mathcal{N}_{2 \, \pm 1}} {{\overline{\zeta}^3}} (1-\mathbf{\mathbf{z}})\sqrt{1-\mathbf{\mathbf{z}}^2} \Bigg[ 3 \mathbf{i} a^6 \mathbf{z}^3 (\mathbf{z}+2)-3 a^5 (\mathbf{z}+1)\left(M (5 \mathbf{z}^2-\mathbf{z}+2)-6 r \mathbf{z}^2\right) \notag \\
        & +\mathbf{i} a^4 \left(2 M^2 (2 \mathbf{z}^4-2 \mathbf{z}^3-6 \mathbf{z}^2-3 \mathbf{z}-3)-3 M r (6 \mathbf{z}^4-17 \mathbf{z}^2-10 \mathbf{z}-3)+9 r^2 \mathbf{z} (\mathbf{z}^3-4 \mathbf{z}-2)\right) \notag \\
        & +a^3 \Big(4 M^3 (\mathbf{z}^3+2 \mathbf{z}^2+2 \mathbf{z}+1)+6 M^2 r (2 \mathbf{z}^4+4 \mathbf{z}^3-2 \mathbf{z}^2-4 \mathbf{z}-1) \notag \\
        &-3 M r^2 (6 \mathbf{z}^4+21 \mathbf{z}^3-17 \mathbf{z}-4)+6 r^3 (\mathbf{z}^4+5 \mathbf{z}^3-5 \mathbf{z}-1)\Big) \notag \\
        & -\mathbf{i} a^2 r \Big(4 M^3 \mathbf{z} (\mathbf{z}+2)+6 M^2 r (5 \mathbf{z}^3+8\mathbf{z}^2+\mathbf{z}-1)-3 M r^2 (16 \mathbf{z}^3+29 \mathbf{z}^2+2 \mathbf{z}-5) \notag \\
        &+9 r^3 (2 \mathbf{z}^3+4 \mathbf{z}^2-1)\Big) \notag \\
        & +2 a r^2 \Big(2 M^3 (2 \mathbf{z}+1)-M^2 r (15 \mathbf{z}^2+14 \mathbf{z}+1)+3 M r^2 \mathbf{z} (8 \mathbf{z}+7)-9 r^3 \mathbf{z} (\mathbf{z}+1)\Big) \notag \\
        &-3 \mathbf{i} r^5 (2 M-r) (2 \mathbf{z}+1) \Bigg] \,,\\
        \hat{h}_{vr}^{(\pm 1 )} ~ & = ~\hat{h}_{rr}^{(\pm 1 )} ~ = ~ \hat{h}_{rz}^{(\pm 1)} ~ = ~ \hat{h}_{r\psi}^{(\pm 1)} = 0\,, \\
        \hat{h}_{zz}^{(\pm 1)} & = - \mathbf{i} \mathcal{N}_{2 \, \pm1} \zeta \sqrt{\frac{1-\mathbf{z}}{1+\mathbf{z}}} \Bigg[3 a^3-9 \mathbf{i} a^2 \mathbf{z} (M-r)+a \left(M^2 (6 \mathbf{z}+4)-9 M (2 r \mathbf{z}+r)+9 r^2 \mathbf{z}\right) \notag \\
        & -6 \mathbf{i} M^2 r+3 \mathbf{i} r^3\Bigg]\,,\\
        \hat{h}_{z\psi}^{(\pm 1)} & = \mp \frac{\mathcal{N}_{2 \, \pm 1}} {{\overline{\zeta}^3}} (1-\mathbf{\mathbf{z}})\sqrt{1-\mathbf{\mathbf{z}}^2} \Bigg[ 3 \mathbf{i} a^6 \mathbf{z}^2 (\mathbf{z}+1)-3 a^5 \mathbf{z} (3 M \mathbf{z}-r (3 \mathbf{z}+2)) \notag \\
        & +\mathbf{i} a^4 \left(2 M^2 \mathbf{z}^2 (2 \mathbf{z}-1)-6 M r (3 \mathbf{z}^3-2 \mathbf{z}-1)+3 r^2 (3 \mathbf{z}^3-3 \mathbf{z}-1)\right) \notag \\
        & +a^3 r \left(2 M^2 (6 \mathbf{z}^3+6 \mathbf{z}^2-2 \mathbf{z}-3)-9 M r (2 \mathbf{z}^3+3 \mathbf{z}^2-1)+3 r^2 (2 \mathbf{z}^3+5 \mathbf{z}^2-1)\right) \notag \\
        & -\mathbf{i} a^2 r \left(M^3 (8 \mathbf{z}+4)+2 M^2 r (3 \mathbf{z}^2+6 \mathbf{z}-1)-18 M r^2 \mathbf{z} (\mathbf{z}+1)+3 r^3 \mathbf{z} (3 \mathbf{z}+1)\right) \notag \\
        & -a r^3 \left(4 M^2 (3 \mathbf{z}+1)-3 M r (2 \mathbf{z}+1)-3 r^2\right)+3 \mathbf{i} r^4 \left(2 M^2-r^2\right)\Bigg] \,,\\
        \hat{h}_{\psi\psi}^{(\pm 1)} & = \frac{\mathbf{i} \mathcal{N}_{2 \, \pm1}}{\zeta \overline{\zeta}^2} \sqrt{1 -\mathbf{z}^2}\Bigg[ (1+\mathbf{z}) (1-\mathbf{z})^2 (a^2+r^2)^2 \Bigg(3 a^3-9 \mathbf{i} a^2 \mathbf{z} (M-r) \notag \\
        & +a \left(M^2 (6 \mathbf{z}+4)-9 M (2 r \mathbf{z}+r)+9 r^2 \mathbf{z}\right)-6 \mathbf{i} M^2 r+3 \mathbf{i} r^3\Bigg) \notag \\
        & +a^2 (1-\mathbf{z}^2)^2 \Bigg(3 a^4+6 \mathbf{i} a^3 (M-r)-2 a^2 M^2-2 \mathbf{i} a (M-r) (2 M^2+6 M r-3 r^2) \notag 
        \\
        & -3 r^2 (2M-r)^2\Bigg) \left(a (\mathbf{z}^2+\mathbf{z}+1)+3 \mathbf{i} r \mathbf{z}\right) \Bigg] \notag \\
        & + 2 a \frac{\mathcal{N}_{2,\pm1}}{\zeta \overline{\zeta}^3} \sqrt{1 -\mathbf{z}^2} (1-\mathbf{z})^2 (1+\mathbf{z}) (a^2+r^2)\Bigg[ 3 a^5 \mathbf{z} (\mathbf{z}^2+\mathbf{z}+1)-3 \mathbf{i} a^4 (2 r \mathbf{z}+r) \notag \\
        &+a^3 \left(4 M^2 \mathbf{z} (\mathbf{z}^2+\mathbf{z}+1)-3 r (2 M-r) (3 \mathbf{z}^3+3 \mathbf{z}^2+\mathbf{z}-1) \right) \notag \\
        &- 2 \mathbf{i} a^2 r \left(M^2 (6 \mathbf{z}^3+6 \mathbf{z}^2+4 \mathbf{z}-1)-9 M r \mathbf{z} (\mathbf{z}^2+\mathbf{z}+1)+3 r^2 \mathbf{z} (\mathbf{z}^2+\mathbf{z}+1)\right) \notag \\
        & -a r (4 M^3+4 M^2 r - 3 r^3)(2 \mathbf{z}+1) +3 \mathbf{i} r^4 (2M-r) (2 \mathbf{z}+1) \Bigg] \,,
    \end{align}
\end{subequations}
where we defined the complex function of the angle $\psi$
\begin{equation} 
    \label{eq:N_2,pm1}
    \mathcal{N}_{2 \, \pm1}\equiv \frac{c_{2 \, \pm1}}{48}\sqrt{\frac{5}{\pi}} e^{\mathbf{i} \psi}
\end{equation}
with
\begin{equation}
\label{eq:bolds_z_and_i_pm1}
\mathbf{z} \equiv
\begin{cases}
    z &\mbox{if } m=+1 \\
    -z & \mbox{if } m=-1
\end{cases} \,,
\quad \mbox{and} \quad
\mathbf{i}\equiv
\begin{cases}
    i &\mbox{if  } m=+1 \\
    -i & \mbox{if  } m=-1
\end{cases} \,.
\end{equation}

\subsection{\texorpdfstring{$m=\pm2$}{m=±2} modes}
Finally, the $m=\pm2$ components, which carry the highest azimuthal dependence, are given explicitly below.
\begin{subequations} 
    \label{eq:hmunu_m_pm2}
    \begin{align}
        \hat{h}_{vv}^{(\pm 2)} & = \frac{\mathcal{N}_{2 \, \pm2}}{\overline{\zeta}^3} (\mathbf{z}^2-1) \Bigg[ 9 \mathbf{i} a^5 \mathbf{z}^3-3 a^4 \left(M (6 \mathbf{z}^3+6 \mathbf{z}^2-2 \mathbf{z}+2)-3 r \mathbf{z}^2 (2 \mathbf{z}+3)\right) \notag \\
        & -3 \mathbf{i} a^3 \left(2 M^2 (\mathbf{z}+1) (\mathbf{z}^2+\mathbf{z}-4)-6 M r (\mathbf{z}^3+5 \mathbf{z}^2+\mathbf{z}+1)+3 r^2 \mathbf{z} (\mathbf{z}^2+6 \mathbf{z}+3)\right) \notag \\
        & -a^2 \left(8 M^3 (\mathbf{z}+2)+6 M^2 r (\mathbf{z}+1) (7 \mathbf{z}-5)-36 M r^2 (7 \mathbf{z}^2+2\mathbf{z}-5)+9 r^3 (3 \mathbf{z}^2+6 \mathbf{z}+1)\right) \notag \\
        & +3 \mathbf{i} a r \left(8 M^3+4 M^2 r (5 \mathbf{z}+2)-4 M r^2 (7 \mathbf{z}+5)+3 r^3 (3 \mathbf{z}+2)\right) \notag \\
        & +9 r^3 (2M - r)^2 \Bigg]\,, \\
        \hat{h}_{vz}^{(\pm 2)} & = \pm \mathbf{i} \frac{\mathcal{N}_{2 \, \pm2}}{\overline{\zeta}^2} (1-\mathbf{z}) \Bigg [ 3 a^5 \mathbf{z}^2 (3 \mathbf{z}+1)-6 \mathbf{i} a^4 \mathbf{z} (-3 M \mathbf{z} (\mathbf{z}+1)+3 r \mathbf{z} (\mathbf{z}+1)+r) \notag \\
        & -a^3 \left(2 M^2 \mathbf{z}^2 (3 \mathbf{z}+5)-3 (3 \mathbf{z}^3+9 \mathbf{z}^2+3 \mathbf{z}+1) (2 M r-r^2)\right) \notag \\
        & +2 \mathbf{i} a^2 r \left(4 M^2 (3 \mathbf{z}^2+\mathbf{z}-3)-9 M r (2 \mathbf{z}^2+\mathbf{z}+1)+6 r^2 \mathbf{z}^2\right) \notag \\
        & +a r \left(16 M^3+2 M^2 r (3 \mathbf{z}+5)-3 r^3 (\mathbf{z}+3)\right)-6 \mathbf{i} r^4 (2 M-r)\Bigg]\,,\\
        \hat{h}_{v\psi}^{(\pm 2)} & = 2 \frac{  { \mathcal{N}_{2 \, \pm2}}  }{\overline{\zeta}^3} (1-\mathbf{z})^2(1+\mathbf{z}) \Bigg [ 3 \mathbf{i} a^6 z^3-3 a^5 (z+1) \left(M (3 z-1) z+M-3 r z^2\right) \notag \\
        & +\mathbf{i} a^4 \left(M^2 (-4 z^3+3 z^2+21 z+12)+3 M r (6 z^3+12 z^2+5 z+3)-9 r^2 z (z^2+3 z+1)\right) \notag \\
        & -a^3 \Big(4 M^3 (z^2+3 z+2)+3 M^2 r (2 z^3+6 z^2+z-5) \notag \\
        &-3 M r^2 (3 z^3+18 z^2+18 z+5) + 3 r^3 (z^3+9 z^2+9 z+1)\Big) \notag \\
        &+ \mathbf{i} a^2 r \left(4 M^3 (z+3)+3 M^2 r (5 z^2+11 z+8)-3 M r^2 (8 z^2+21 z+7)+9 r^3 (z^2+3 z+1)\right) \notag \\
        &-a r^2 \left(8 M^3 - M^2 r (15 z+13)+6 M r^2 (4 z+3)-9 r^3 (z+1)\right)+3 \mathbf{i} r^5 (2 M-r) \Bigg] \,,\\
        \hat{h}_{vr}^{(\pm 2)} ~ & = ~ \hat{h}_{rr}^{(\pm 2)} ~ = ~ \hat{h}_{rz}^{(\pm 2)} ~ = ~ \hat{h}_{r\psi}^{(\pm 2)} ~ = ~ 0\,, \\
        \hat{h}_{zz}^{(\pm 2)} & = -\mathbf{i} \mathcal{N}_{2 \, \pm2} \zeta \left(\frac{1-\mathbf{z}}{1+\mathbf{z}}\right) \Bigg[ a^3 (9 \mathbf{z}+6)+9 \mathbf{i} a^2 \left(2 M (\mathbf{z}+1)-r (2 \mathbf{z}+1)\right) \notag \\
        &-a \left(2 M^2 (3 \mathbf{z}+4)-18 M r (\mathbf{z}+1)+9 r^2 \mathbf{z}\right)+3 \mathbf{i} r (2M^2-r^2) \Bigg] \,,\\
        \hat{h}_{z\psi}^{(\pm 2)} & = \mp \mathbf{i} \frac{\mathcal{N}_{2 \, \pm2}}{\overline{\zeta}^2}(1-\mathbf{z}^2) \Bigg[ 3 a^6 \mathbf{z}^2 (2 \mathbf{z}+1)-6 \mathbf{i} a^5 \mathbf{z} (-3 M \mathbf{z} (\mathbf{z}+1)+3 r \mathbf{z} (\mathbf{z}+1)+r) \notag \\
        & -a^4 \left(+2 M^2 \mathbf{z}^2 (4 \mathbf{z}+5)-6 M r (6\mathbf{z}^3+9\mathbf{z}^2+4\mathbf{z}+1)+3 r^2 (6 \mathbf{z}^3+12 \mathbf{z}^2+6 \mathbf{z}+1)\right) \notag \\
        & +2 \mathbf{i} a^3 r \left(2 M^2 (3\mathbf{z}^3 +6\mathbf{z}^2-4\mathbf{z} -6)-9 M r (\mathbf{z}+1)^3+3 r^2 (\mathbf{z}^3+5 \mathbf{z}^2+3 \mathbf{z}+1)\right) \notag \\
        & +a^2 r \left(16 M^3 (\mathbf{z}+1)+2 M^2 r (3 \mathbf{z}^2+12 \mathbf{z}+5)-18 M r^2 (\mathbf{z}+1)^2+3 r^3 \mathbf{z} (3 \mathbf{z}+2)\right) \notag \\
        & -2 \mathbf{i} a r^3 \left(2 M^2 (3 \mathbf{z}+2)-3 M r (\mathbf{z}+1)-3 r^2\right)-6 M^2 r^4+3 r^6\Bigg] \,,\\
        \hat{h}_{\psi\psi}^{(\pm 2)} & = \frac{\mathcal{N}_{2 \, \pm2}}{\zeta \overline{\zeta}^2} (1-\mathbf{z}^2) \Bigg[ \mathbf{i} a^2 (\mathbf{z}^2-1)^2 (a (\mathbf{z}+2)+3 \mathbf{i} r) \Big(3 a^4+12 \mathbf{i} a^3 (M-r) \notag \\
        &+2 a^2 (4 M^2+18 M r-9 r^2) +4 \mathbf{i} a (M-r) (2 M^2+6 M r-3r^2)+3 r^2 (2M - r)^2\Big) \notag \\
        & +(1-\mathbf{z}^2)^2 (a^2+r^2)^2 \Bigg(3 \mathbf{i} a^3 (3 \mathbf{z}+2)- 9 a^2 \left(2 M (\mathbf{z}+1)-r (2 \mathbf{z}+1)\right) \notag \\
        & - \mathbf{i} a \left(M^2 (6\mathbf{z}+8)-18 M r (\mathbf{z}+1)+9 r^2 \mathbf{z}\right)-6 M^2 r + 3 r^3\Bigg)\Bigg] \notag \\
        &+4 \frac{\mathcal{N}_{2,\pm2}}{\zeta \overline{\zeta}^3} (1-\mathbf{z}) (1-\mathbf{z}^2)^2  (a^2+r^2) a \Bigg[3 a^5 \mathbf{z} (\mathbf{z}+1)+3 \mathbf{i} a^4 \left(3 M \mathbf{z} (\mathbf{z}+1)-3 r \mathbf{z} (\mathbf{z}+1)+r \right) \notag \\
        &-a^3 \left(4 M^2 \mathbf{z} (\mathbf{z}+1)-6 M r (3 \mathbf{z} (\mathbf{z}+1)+2)+3 r^2 (3 \mathbf{z}^2 + 3\mathbf{z} +2)\right) \notag \\
        &+\mathbf{i} a^2 r \left(M^2 (6 \mathbf{z}^2 + 6\mathbf{z}-8)-9 M r (\mathbf{z}^2+\mathbf{z}+2)+3 r^2 \mathbf{z} (\mathbf{z}+1)\right) \notag \\
        & +2 a r \left(4 M^3+4 M^2 r-3 r^3\right)-3 \mathbf{i} r^4 (2 M-r) \Bigg] \,,
    \end{align}
\end{subequations}
where, analogously to the case $m\pm 1$, we defined
\begin{equation} 
    \label{eq:N_2pm2}
    \mathcal{N}_{2 \, \pm2}\equiv \frac{c_{2 \, \pm2}}{96}\sqrt{\frac{5}{\pi}} e^{2 \mathbf{i} \psi}
\end{equation}
with
\begin{equation}
\label{eq:bolds_z_and_i_pm2}
\mathbf{z} \equiv
\begin{cases}
    z &\mbox{if } m=+2 \\
    -z & \mbox{if } m=-2
\end{cases} \,,
\quad \mbox{and} \quad
\mathbf{i}\equiv
\begin{cases}
    i &\mbox{if  } m=+2 \\
    -i & \mbox{if  } m=-2
\end{cases} \,.
\end{equation}

\vspace{20pt}

The complete perturbation $h_{\mu\nu}^{\rm ORG}$, obtained by summing the five
contributions above and taking the real part of the result, satisfies the ORG
conditions. One can further verify that, in the zero-spin limit $a=0$, it
reduces to the known quadrupolar tidal perturbation of the Schwarzschild metric
derived in Ref.~\cite{Binnington:2009bb}. Moreover, by expanding the
reconstructed metric to first order in the spin parameter $a$, the associated
Weyl scalars $\psi_0$ and $\psi_4$ agree with those obtained in
Refs.~\cite{Poisson:2014gka, Landry:2015zfa}.

Finally, we have explicitly verified that each mode of the metric perturbation
$\hat{h}_{\mu\nu}^{(m)\,\rm ORG}$ satisfies the linearized Einstein equations. For completeness, we recall that the full expression for the components of the metric perturbation is provided in electronic form as a \texttt{Mathematica} notebook in the supplementary material to this paper.

By performing a suitable gauge transformation from the ORG to the IRG and
rotating from the HH to the Kinnersley tetrad, the tidally perturbed Kerr metric
constructed here can be related to those obtained in
Refs.~\cite{LeTiec:2020bos, Yunes:2005ve}.
We also note that explicit reconstruction formulas in both the IRG and ORG
within the Kinnersley tetrad were recently presented in
Ref.~\cite{Berens:2024czo}.\footnote{See Ref.~\cite{Berens:2025kkm} for an extension of this formalism to Kerr black holes embedded in a spacetime with a nonzero cosmological constant.} Those expressions, however, were neither specialized
to tidal perturbations nor given in the fully explicit analytic form derived here, which facilitates direct use in astrophysical applications (see Secs.~\ref{sec:Secular_Dynamics} and~\ref{sec:ISCO_LR_shifts}). 


\subsection{\texorpdfstring{Fixing the coefficients $c_{2m}$}{Fixing the coefficients c2m}}

We now determine the coefficients $c_{2m}$ entering the metric through the
normalization factors $\mathcal{N}_{2\,0}$, $\mathcal{N}_{2\,\pm1}$, and
$\mathcal{N}_{2\,\pm2}$ in Eqs.~\eqref{eq:N_20}, \eqref{eq:N_2,pm1}, and
\eqref{eq:N_2pm2}.

As anticipated in Sec.~\ref{subsec:HertzPotential}, we use the results of the
standard asymptotic-matching procedure employed in
Refs.~\cite{Poisson:2009qj, Binnington:2009bb, Poisson:2004cw,
Chatziioannou:2012gq} to determine the coefficients $c_{2m}$ by taking the zero-spin limit ($a=0$) of the metric
perturbation~\eqref{eq:ORG} and requiring that it reproduces the quadrupolar order
of the tidally deformed metric derived in Ref.~\cite{Binnington:2009bb} for a
Schwarzschild black hole in the same gauge. This fixes the coefficients $c_{2m}$ in terms of the external tidal moments.

Introducing the complex combination of the tidal tensors
$\mathcal{C}_{ab} \equiv \mathcal{E}_{ab} - i \mathcal{B}_{ab}$, the coefficients read
\begin{equation}
\label{eq:alpha_and_beta_coeff}
\begin{aligned}
c_{2\,0} &=
-4 \sqrt{\frac{2\pi}{15}}\,
\bigl(\mathcal{C}_{11} + \mathcal{C}_{22}\bigr) ,
\\[0.4em]
c_{2\,\pm 1} &=
\mp \frac{8}{3} \sqrt{\frac{\pi}{5}}\,
\bigl(\mathcal{C}_{13} \mp i\,\mathcal{C}_{23}\bigr) ,
\\[0.4em]
c_{2\,\pm 2} &=
\frac{4}{3} \sqrt{\frac{\pi}{5}}\,
\bigl(\mathcal{C}_{11} - \mathcal{C}_{22} \mp 2 i\,\mathcal{C}_{12}\bigr) .
\end{aligned}
\end{equation}
As anticipated in Sec.~\ref{sec:MetricReconstruction}, once the relation between
the Weyl scalar $\psi_0$ and the Hertz potential $\Psi'_{\rm H}$ is taken into
account, these coefficients are consistent with those obtained in
Refs.~\cite{Poisson:2004cw, Chatziioannou:2012gq}.

Finally, for later convenience, we introduce the scalar tidal potentials~\cite{Poisson:2014gka}
\begin{equation}
\label{eq:Eq_and_Bq}
\mathcal{E}^{\rm q} \equiv \mathcal{E}_{ab}\,\Omega^a \Omega^b ,
\qquad
\mathcal{B}^{\rm q} \equiv 
\mathcal{B}_{ab}\,\Omega^a \Omega^b ,
\end{equation}
where $\Omega^a \equiv (\sqrt{1-z^2}\cos\psi,\,
\sqrt{1-z^2}\sin\psi,\,
z)$ is the unit spatial vector.


\section{Secular dynamics in tidally deformed Kerr spacetime}
\label{sec:Secular_Dynamics}

A compelling and not yet fully explored question concerns how self-gravitating systems respond dynamically to external tidal fields. While tidal Love numbers
have been extensively studied for compact objects in a wide range of theoretical
frameworks~\cite{Chakraborty:2025wvs, Chia:2020yla, LeTiec:2020bos,
LeTiec:2020spy, Pani:2015hfa, Gurlebeck:2015xpa, Poisson:2014gka, Kol:2011vg,
Damour:2009va, Landry:2015zfa, Pani:2015nua, Binnington:2009bb,
Chakrabarti:2013lua, Damour:2009vw, Chatziioannou:2020pqz, Flanagan:2007ix,
Hinderer:2007mb, Cardoso:2017cfl, Pereniguez:2025jxq, Charalambous:2021mea,
Perry:2023wmm}, comparatively little attention has been devoted to the dynamics
of matter orbiting a tidally distorted compact object.

Notable exceptions include studies of EMRI systems subject to external
perturbations, which have shown modifications to orbital evolution and the
possible onset of tidal resonances~\cite{Yang:2017aht, Yang:2019iqa,
Bonga:2019ycj, Gupta:2021cno, Gupta:2022fbe}.

Using the expression for the tidally deformed metric derived in
Sec.~\ref{sec:explicit_Kerr_perturbations}, we investigate the motion of a test
particle around a rotating black hole subject to an external tidal
deformation. Our focus is on \emph{secular effects}, namely corrections that
accumulate over many orbital cycles, or equivalently on timescales much longer
than the orbital period. As already mentioned in the Introduction, such effects
are especially relevant for EMRIs, in which a small body orbits a much more
massive black hole for a very large number of cycles.

In this regime, the dynamics is naturally characterized by a separation of
timescales between the fast orbital motion and the much slower evolution induced
by weak external perturbations. This hierarchy provides the basis for a
description in terms of quantities averaged over the orbital period~\cite{Mino:2003yg, Hinderer:2008dm}.

The presence of an external tidal field introduces additional long-term modifications to the orbital dynamics. Characterizing how such tidal deformations affect the secular behavior of orbital observables is therefore crucial, since even small corrections can accumulate over the inspiral and lead
to potentially observable imprints in the phase of the emitted gravitational waves.


\subsection{Secular Hamiltonian}
We begin by examining the case of a test particle of mass $m$ moving on a
timelike geodesic in the tidally deformed spacetime with metric $g_{\mu\nu}$.
To describe the dynamics of this particle, we use the geodesic equation,
including corrections to the Kerr geometry up to linear order in the metric
perturbation $h_{\mu\nu}$. This approximation is valid provided that
self-force effects can be neglected, namely in the limit $m/M \ll 1$, where
$M$ denotes the mass of the Kerr black hole~\cite{Fujita:2016igj}.
We also require $M/\mathcal{R} \ll 1$, where $\mathcal{R}$ is the length scale associated with the external tidal field, and we retain contributions only up to leading quadrupolar order, $\mathcal{O}\!\left((M/\mathcal{R})^2\right)$.

In the secular regime, and to first order in the metric perturbation
$h_{\mu\nu}$, the physical orbit followed by the particle in the perturbed
spacetime $g_{\mu\nu} = \tilde{g}_{\mu\nu} + h_{\mu\nu}$, which we denote by
$\Gamma'$, can be accurately approximated by an averaged geodesic $\Gamma$,
defined within the same perturbed spacetime. This averaged geodesic can be
interpreted as a \emph{secular orbit} in the tidally deformed background.

For bound geodesic motion in Kerr spacetime, the radial, polar, and azimuthal motions are characterized by three independent fundamental frequencies. It is therefore convenient to parametrize the orbit using three phase variables
$(q_r, q_z, q_\psi)$, which label the oscillations associated with the radial,
polar, and azimuthal motions, respectively, and evolve linearly with respect to
a suitable parameter along the geodesic~\cite{Hinderer:2008dm,
vandeMeent:2019cam, Schmidt:2002qk, Drasco:2003ky}.

For a generic, non-resonant orbit, the secular average of a quantity
$\mathcal{A}$ is defined as~\cite{Drasco:2005is}
\begin{equation}
\label{eq:secularaverage_general}
\langle \mathcal{A} \rangle
= \frac{1}{(2\pi)^3}
  \int_0^{2\pi} dq_\psi
  \int_0^{2\pi} dq_z
  \int_0^{2\pi} dq_r \,
  \mathcal{A}(q_r,q_z,q_\psi)\big|_{\Gamma} \, .
\end{equation}
We stress that the secular averaging procedure adopted here is valid for non-resonant orbits, namely when the fundamental frequencies associated with the $r$, $z$, and $\psi$ motions are incommensurate. Near orbital resonances, combinations of these frequencies can become commensurate, allowing terms that are normally oscillatory to accumulate secularly over long timescales, leading to a breakdown of the averaged description~\cite{Hinderer:2008dm,Flanagan:2010cd,Fujita:2016igj,Lynch:2024ohd}. Throughout this work, we therefore restrict our attention to non-resonant configurations.

As already mentioned, at the order considered here, the secularly averaged
geodesic $\Gamma$ coincides with the physical orbit $\Gamma'$. More precisely,
differences arise only at higher order, for example when including first-order corrections in the mass ratio $m/M$ or second-order contributions in the metric perturbation $h_{\mu\nu}$~\cite{Fujita:2016igj, Yang:2017aht,
Camilloni:2023rra}.

Moreover, to first order in the tidal perturbation, the secularly averaged
energy $E$ and angular momentum $L$, both defined per unit rest mass, remain
conserved along the secular orbit $\Gamma$~\cite{Fujita:2016igj,
Camilloni:2023rra}. We stress that $E$ and $L$ here denote secularly averaged quantities. While the perturbed spacetime $g_{\mu\nu}$ remains stationary and thus admits a conserved energy along exact geodesics, the axial symmetry is broken by the perturbation $h_{\mu\nu}$, so that $L$ is not conserved instantaneously. However, the secular averaging in Eq.~\eqref{eq:secularaverage_general} effectively restores axisymmetry, leading to conservation of the averaged angular momentum.

The Hamiltonian of the test particle is given by
\begin{equation}
\label{eq:hamiltonian_def}
H = \frac{1}{2}\, g_{\mu\nu} u^\mu u^\nu \, .
\end{equation}
Expanding to first order in the metric perturbation $h_{\mu\nu}$, we write
\begin{equation}
H \simeq
\frac{1}{2}\, \tilde{g}_{\mu\nu}\tilde{u}^\mu \tilde{u}^\nu
+ \tilde{g}_{\mu\nu}\tilde{u}^\mu u_{(1)}^\nu
+ \frac{1}{2}\, h_{\mu\nu}\tilde{u}^\mu \tilde{u}^\nu \, ,
\end{equation}
where we used that the four-velocity admits the expansion $u^\mu \simeq \tilde{u}^\mu + u_{(1)}^\mu$.
Here, $\tilde{u}^\mu$ denotes the four-velocity along the geodesic of the unperturbed Kerr spacetime, while $u_{(1)}^\mu$ represents the correction induced by the tidal perturbation $h_{\mu\nu}$.

We now restrict our analysis to circular equatorial orbits ($z=0$). In this case, both the radial and polar motions are absent, and the only nonvanishing
components of the four-velocity are $u_v$ and $u_\psi$, which can be expressed
in terms of the conserved energy $E$ and angular momentum $L$.

Accordingly, the general secular average defined in
Eq.~\eqref{eq:secularaverage_general} simplifies to an average over the azimuthal angle,
\begin{equation}
\label{eq:secularaverage_circular}
\langle \mathcal{A} \rangle
= \frac{1}{2\pi}
  \int_0^{2\pi} \mathcal{A}(\psi)\big|_{\Gamma}\, d\psi \, ,
\end{equation}
since the dependence on the radial and polar phases is absent, and the phase
variable $q_\psi$ coincides with the azimuthal coordinate $\psi$ for circular
equatorial motion.

As indicated in Eq.~\eqref{eq:ORG}, the metric perturbation is given by twice the
real part of the sum over modes with $m\in[-2,2]$. However, in the equatorial
plane, after performing the secular average, only the $m=0$ component
contributes, since the azimuthal averaging over a circular equatorial orbit
eliminates all non-axisymmetric ($m\neq 0$) terms. Using the results of
Sec.~\ref{sec:explicit_Kerr_perturbations}, we obtain%
\footnote{This averaging does not affect the background metric
$\tilde{g}_{\mu\nu}$, owing to the axial symmetry of the Kerr spacetime.}

\begin{subequations}
    \label{eq:hmunuave}
    \begin{align}
    \langle h_{vv}\rangle ~=&~ -\frac{\langle \mathcal{E}^{\rm q} \rangle}{r^3}  \left(2 a^4 M - a^2 r (2 M^2+4 M r-3 r^2)+r^3 (r-2 M)^2\right) \,, \\
    \langle h_{vz}\rangle ~=&~ a \frac{\langle \mathcal{B}^{\rm q} \rangle}{r} \left(a^2 (r-2 M)+2 M^2 r-r^3\right) \,,  \\
    \langle h_{v\psi}\rangle ~=&~ 2 a \frac{\langle \mathcal{E}^{\rm q} \rangle}{r^3} \left(a^4 M - a^2 r (M^2+Mr-r^2)+r^3 (M-r)^2\right) \,, \\
    \langle h_{zz}\rangle ~=&~ \langle \mathcal{E}^{\rm q} \rangle r \left(a^2 (2 M-r)-2 M^2 r+r^3\right) \,, \\
    \langle h_{z\psi}\rangle ~=&~  \frac{\langle \mathcal{B}^{\rm q} \rangle}{r}  \left(a^2+r^2\right) \left(a^2 (2 M-r)-2 M^2 r+r^3\right) \,,     \\
    \langle  h_{\psi\psi}\rangle ~=&~ -\frac{\langle \mathcal{E}^{\rm q} \rangle}{r^3} \left(2 a^6 M+a^4 (r^3-2 M^2 r) - 2 a^2 r^4 (M - r) - 2 M^2 r^5 + r^7 \right) \,, \\
    \langle h_{vr}\rangle~=&~\langle h_{rr}\rangle ~=\langle h_{rz}\rangle = \langle h_{r\psi}\rangle = 0~,
    \end{align}
\end{subequations}
where, using Eq.~\eqref{eq:Eq_and_Bq}, the electric tidal scalar $\mathcal{E}^{\rm q}$ is written explicitly as
\begin{subequations}
    \begin{align}
    \label{eq:scalarEq}
   \mathcal{E}^{\rm q}  & =   \frac{\mathcal{E}_{11}-\mathcal{E}_{22}}{2}  \left(1-z^2\right) \cos 2\psi   + \frac{\mathcal{E}_{11}+\mathcal{E}_{22}}{2} \left(1-3 z^2\right)
   + \mathcal{E}_{12} \left(1-z^2\right)  \sin 2 \psi \notag \\
   & + 2 z  \sqrt{1-z^2} \left(\mathcal{E}_{13} \cos\psi + \mathcal{E}_{23} \sin \psi \right) \,,
   \end{align}
\end{subequations}
and its secular average on the equatorial plane  is given by 
$\langle \mathcal{E}^{\rm q }\rangle \equiv (\mathcal{E}_{11}+\mathcal{E}_{22})/2$. 
Equivalently, one obtains an expression analogous to \eqref{eq:scalarEq} for the magnetic tidal scalar $\mathcal{B}^{\rm q}$, and its secular average on the equatorial plane is given by $\langle \mathcal{B}^{\rm q} \rangle \equiv (\mathcal{B}_{11}+\mathcal{B}_{22})/2$. 

Performing the secular average of the Hamiltonian to first order in the tidal
perturbation yields
\begin{equation}
\label{eq:hamiltonian_averaged}
\langle H \rangle
= \langle H \rangle_0 + \eta\, \langle H \rangle_1 \, ,
\end{equation}
where
\begin{equation}
\label{eq:eta}
\eta \equiv - \frac{M^2}{2}\, \langle \mathcal{E}^{\rm q} \rangle
\end{equation}
is a dimensionless perturbative parameter encoding information about the tidal
environment~\cite{Yang:2017aht, Camilloni:2023rra, Grilli:2024fds}. For completeness, and to assess the regime of validity of our perturbative treatment, we provide analytical and numerical estimates of the magnitude of $\eta$ in App.~\ref{app:eta_estimate}.
We note that, although some components of the metric perturbation depend on the
magnetic tidal scalar $\mathcal{B}^{\rm q}$, even after the averaging procedure,
these components do not contribute to the Hamiltonian. As a result, any
dependence on $\langle \mathcal{B}^{\rm q} \rangle$ drops out.

The unperturbed averaged Hamiltonian reads
\begin{equation}
\label{eq:H0_averaged}
\langle H \rangle_0
= -\frac{1}{2 r \Delta}
\left[
\langle E^2 \rangle \bigl(r^3 + a^2(r + 2 M)\bigr)
- \langle L^2 \rangle (r-2 M)
- 4 a M \langle E \rangle \langle L \rangle
\right] ,
\end{equation}
while the first-order correction in $\eta$ is given by 
\begin{equation}
\label{eq:H1_averaged}
\begin{split}
\langle H \rangle_1
& = - \frac{1}{M^2 r^3}
\Big[
\langle E^2 \rangle \bigl(r^5 + a^2 r(3r^2 - 2 M^2) + 2a^4 M\bigr) + \langle L^2 \rangle \bigl(r^3 - 2r M^2 + 2a^2 M\bigr) \\
& - 4 a \langle E \rangle \langle L \rangle \bigl(r^3 - r M^2 + a^2 M\bigr)
\Big] .
\end{split}
\end{equation}
In the zero-spin limit $a=0$, Eqs.~\eqref{eq:H0_averaged} and
\eqref{eq:H1_averaged} reproduce the results obtained in
Refs.~\cite{Yang:2017aht, Camilloni:2023rra} for a tidally deformed
Schwarzschild spacetime.

The secular Hamiltonian~\eqref{eq:hamiltonian_averaged} provides a powerful and
convenient framework to analyze long-term tidal effects on the location and properties of both the ISCO and the LR. These two orbits play a central role in black-hole physics, as they
encode essential information about the behavior of matter, radiation, and gravitational perturbations in the SF region. In the next section, we examine in detail how tidal fields modify the properties of the ISCO and the LR.


\section{Tidal deformation of ISCO and LR} 
\label{sec:ISCO_LR_shifts}

In this section, we compute how the location and properties of the ISCO and of the LR of a Kerr black hole are modified by tidal effects, and we investigate how these modifications depend on the black-hole spin.
As anticipated in Sec.~\ref{sec:Secular_Dynamics}, all results are obtained to
first order in the tidal parameter $\eta$, using the secular Hamiltonian
$\langle H \rangle$.

The ISCO is not only a mathematically well-defined geodesic of the Kerr
geometry~\cite{Bardeen:1972fi}, but also an object of direct astrophysical
relevance. The binding energy at the ISCO sets the radiative efficiency of thin
accretion disks, and measurements of disk spectra have therefore been used to
infer black-hole spins through continuum-fitting methods~\cite{Narayan:2005ie,
Zhang:1997dy}. Similarly, the $\mathrm{K}\alpha$ iron line originates in the innermost regions of the accretion disk, and its profile is sensitive to the location of the disk's inner edge, which is often identified with the
ISCO~\cite{Fabian:1989ej, Fabian:2000nu}.

From a gravitational-wave perspective, the ISCO marks the end of the adiabatic
inspiral and the onset of the plunge phase. In EMRI systems, even small tidal
shifts in the location of the ISCO or in the associated orbital constants can
accumulate over many cycles and lead to potentially detectable dephasings in
the gravitational-wave signal~\cite{Ori:2000zn}.

The LR is likewise of strong physical and observational relevance.
Early studies showed that the LR controls the optical appearance of a collapsing
star and the exponential fade-out of its observed luminosity~\cite{Ames:1968apj}.
More recently, the LR has been recognized as playing a pivotal role in a variety
of observational and theoretical contexts, including the determination of the
photon-ring structure and, consequently, of the black-hole shadow~\cite{Falcke:1999pj}.
The angular frequency and instability timescale (i.e., Lyapunov exponent) of the
LR are directly related to the real and imaginary parts of black-hole quasinormal
modes in the eikonal limit~\cite{Goebel:1972apj, Ferrari:1984zz, Mashhoon:1985cya,
Berti:2005eb, Cardoso:2008bp, Cardoso:2021qqu, Yang:2012he}. Furthermore, the LR
also determines the critical impact parameter relevant for high-energy scattering
and threshold phenomena in numerical collision experiments~\cite{Pretorius:2007jn}.

Computing tidal-induced secular shifts of LR quantities, such as the radius,
orbital frequency, impact parameter, and instability rate, is therefore important, as such shifts could leave measurable imprints on black-hole shadows, ringdown
signals, and critical scattering dynamics~\cite{Cardoso:2021qqu}.

At this point, we should note that the tidal perturbations derived in this paper are obtained in a specific gauge, the ORG. When considering their physical consequences, it is therefore important to distinguish which quantities do and which do not depend on this choice of gauge. If we restrict to gauge transformations that are bound along the particle's trajectory, the averaged parts of the actions (notably $E$ and $L$) cannot be changed by gauge transformations (see e.g. Eq.~(4.5) of \cite{Fujita:2016igj}). Gauge invariance, within the same class of gauges, of the averaged orbital frequency $\Omega\equiv \langle u^\psi/u^v \rangle$ and redshift invariant $U \equiv \langle u^v \rangle$ is similarly well established \cite{Sago:2008id,Barack:2011ed,Fujita:2016igj}. However, the shifts in $r$ recorded below are not gauge invariant as they are sensitive to first-order transformations of $r$.


\subsection{ISCO shifts}

We begin by computing the shift induced by the tidal deformation on the location
and properties of the ISCO of a Kerr black hole. We recall that, in the
equatorial plane of the Kerr spacetime, there exist two ISCOs corresponding to
prograde and retrograde orbits, respectively. These orbits can be determined by
imposing the following three conditions~\cite{Isoyama:2014mja}
\begin{equation}
\label{eq:ISCO_cond}
\langle H \rangle \big|_{r=r_{\rm ISCO}} = -\frac{1}{2} \, , \qquad
\frac{d\langle H \rangle}{dr}\bigg|_{r=r_{\rm ISCO}} = 0 \, , \qquad
\frac{d^2 \langle H \rangle}{dr^2}\bigg|_{r=r_{\rm ISCO}} = 0 \, .
\end{equation}
The solution of these equations yields the ISCO radius, as well as the specific
energy and specific angular momentum of a test particle moving on this orbit.
Up to first order in the perturbative parameter $\eta$, we write
\begin{equation}
\label{eq:ISCO_corrections_rEL}
\begin{aligned}
r^{\sigma}_{\rm ISCO} &\simeq
r^{\sigma}_{0\,{\rm ISCO}} + \eta\, r^{\sigma}_{1\,{\rm ISCO}} \, , \\
E^{\sigma}_{\rm ISCO} &\simeq
E^{\sigma}_{0\,{\rm ISCO}} + \eta\, E^{\sigma}_{1\,{\rm ISCO}} \, , \\
L^{\sigma}_{\rm ISCO} &\simeq
L^{\sigma}_{0\,{\rm ISCO}} + \eta\, L^{\sigma}_{1\,{\rm ISCO}} \, ,
\end{aligned}
\end{equation}
where $\sigma=\pm1$ labels prograde ($\sigma=+1$) and retrograde ($\sigma=-1$)
orbits, respectively. 
Analogously, one can compute the orbital frequency $\Omega\equiv \langle u^\psi/u^v \rangle$ and the
redshift invariant $U \equiv \langle u^v \rangle$~\cite{Detweiler:2008ft, Sago:2008id, Shah:2012gu, vandeMeent:2015lxa},%
\footnote{For circular equatorial orbits, these quantities coincide with $\langle u^\phi/u^t \rangle$ and $\langle u^t \rangle$ in Boyer--Lindquist coordinates, since  $(v,\psi)$ differ from $(t,\phi)$  only by functions of the radial coordinate $r$
(cfr.~Eq.~\eqref{eq:ingoing_coordinates}).}
\begin{equation}
\label{eq:ISCO_corrections_freq_red}
\Omega^{\sigma}_{\rm ISCO} \simeq
\Omega^{\sigma}_{0\,{\rm ISCO}} + \eta\, \Omega^{\sigma}_{1\,{\rm ISCO}} \, ,
\qquad
U^{\sigma}_{\rm ISCO} \simeq
U^{\sigma}_{0\,{\rm ISCO}} + \eta\, U^{\sigma}_{1\,{\rm ISCO}} \, .
\end{equation}

By solving Eqs.~\eqref{eq:ISCO_cond} at zeroth order in the tidal parameter
$\eta$, one recovers the well-known expressions for the radial coordinate, the
specific energy, and the specific angular momentum of the unperturbed ISCOs of a Kerr black hole, in agreement with the seminal results of
Refs.~\cite{Chandrasekhar:1985kt, Bardeen:1972fi}:
\begin{equation}
\label{eq:risco_unperturbed}
r^{\sigma}_{0\,{\rm ISCO}}
= M \left[3 + Z_2 - \sigma
\sqrt{(3 - Z_1)(3 + Z_1 + 2 Z_2)} \right]\, ,
\end{equation}
where
\begin{equation}
\label{eq:z1_z2_isco}
Z_1
= 1 + (1 - \chi^2)^{1/3}
\left[(1 - \chi)^{1/3} + (1 + \chi)^{1/3}\right] ,
\qquad
Z_2 = \sqrt{3 \chi^2 + Z_1^2} \, ,
\end{equation}
and $\chi\equiv a/M$ is the dimensionless spin parameter.

The unperturbed energy and angular momentum of the ISCO can be conveniently
expressed in terms of the dimensionless unperturbed ISCO radius $r_0 \equiv r^{\sigma}_{0\,{\rm ISCO}}/M$:
\begin{equation}
\label{eq:Eisco_unperturbed}
E^{\sigma}_{0\,{\rm ISCO}}
= \sqrt{1 - \frac{2}{3 r_0}} \, , \qquad
L^{\sigma}_{0\,{\rm ISCO}}
= M \left[\chi \sqrt{1 - \frac{2}{3 r_0}}
+ \sigma\, \frac{r_0}{\sqrt{3}} \right]\, .
\end{equation}
Applying the same procedure to
Eqs.~\eqref{eq:ISCO_corrections_freq_red} at leading order in $\eta$ yields
\begin{equation}
\label{eq:freq_red_ISCO_unperturbed}
\Omega^{\sigma}_{0\,{\rm ISCO}}
\, M = \frac{\sigma}{r_0^{3/2} + \sigma \chi} \, , \qquad
U^{\sigma}_{0\,{\rm ISCO}}
= \frac{r_0^{3/2} + \sigma \chi}
{r_0^{3/4}
\sqrt{r_0^{1/2}(r_0 - 3) + 2 \sigma \chi}} \, .
\end{equation}
By solving Eqs.~\eqref{eq:ISCO_cond} to first order in the tidal parameter
$\eta$, we can compute the leading perturbative corrections to the quantities
characterizing the ISCO.

For the first-order tidal shift of the ISCO radius, we find
\begin{equation}
\label{eq:risco_perturbed}
r^{\sigma}_{1\,{\rm ISCO}}
= -2 M\frac{r_0^2 - 2 r_0 + \chi^2}{r_0}\,
\frac{\sqrt{r_0}\,\mathcal{A}(\chi,r_0)
+ \sigma\, \mathcal{B}(\chi,r_0)}
{\mathcal{C}(\chi,r_0)
+ \sigma\,\mathcal{D}(\chi,r_0)} \, ,
\end{equation}
where the functions $\mathcal{A}$, $\mathcal{B}$, $\mathcal{C}$, and
$\mathcal{D}$ are given by
\begin{subequations}
\label{eqs:ABCD_risco_perturbed}
\begin{align}
\mathcal{A}(\chi,r_0)
&= \chi^6(15 r_0 - 84)
+ \chi^4(13 r_0^2 + 30 r_0)
+ \chi^2(23 r_0^5 - 106 r_0^4 + 44 r_0^3 + 72 r_0^2)
\nonumber\\
&\quad
+ \bigl(4 r_0^8 - 34 r_0^7 + 101 r_0^6
- 126 r_0^5 + 48 r_0^4\bigr) \, , \\[0.4em]
\mathcal{B}(\chi,r_0)
&= 21 \chi^7
+ \chi^5(93 r_0 - 36 r_0^2)
+ \chi^3(3 r_0^5 + 19 r_0^4 + 9 r_0^3 - 120 r_0^2)
\nonumber\\
&\quad
+ \chi(19 r_0^7 - 106 r_0^6 + 200 r_0^5
- 90 r_0^4 - 12 r_0^3) \, , \\[0.4em]
\mathcal{C}(\chi,r_0)
&= \chi^4\sqrt{r_0}(34 - 9 r_0)
- 2 \chi^2 r_0^{3/2}\bigl[r_0(7 r_0 - 48) + 50\bigr]
\nonumber\\
&\quad
+ r_0^{5/2}\bigl[(r_0 - 6) r_0 (3 r_0 - 16) - 72\bigr] \, , \\[0.4em]
\mathcal{D}(\chi,r_0)
&= -9 \chi^5
- 2 \chi^3 r_0(r_0 + 6)
+ \chi r_0^2\bigl[r_0(39 r_0 - 172) + 156\bigr] \, .
\end{align}
\end{subequations}

For the first-order tidal correction to the ISCO energy, we find
\begin{equation}
\label{eq:Eisco_perturbed}
E^\sigma_{1\,{\rm ISCO}}
=
\frac{
\mathcal{P}_E(\chi,r_0,r_1)
+ \sigma\, \sqrt{r_0}\, \mathcal{Q}_E(\chi,r_0,r_1)
}{
2\, r_0^{11/4}
\left[\sqrt{r_0}(r_0 - 3) + 2 \chi \sigma \right]^{3/2}
} \, ,
\end{equation}
where, for notational convenience, we defined
$r_1 \equiv r^\sigma_{1\,{\rm ISCO}}/M$, and where the functions
$\mathcal{P}_E$ and $\mathcal{Q}_E$ are given by
\begin{subequations}
\label{eqs:PQ_Eisco_perturbed}
\begin{align}
\mathcal{P}_E(\chi,r_0,r_1)
&=
6 \chi^6
+ 6 \chi^4 r_0
- 4 \chi^4 r_0^2
- \chi^2 \left(8 r_0^5 - 32 r_0^4 + 22 r_0^3 + 16 r_0^2 + 3 r_0 r_1 \right)
\nonumber\\
&\quad
- 4 r_0^7
+ 22 r_0^6
- 44 r_0^5
+ 32 r_0^4
+ r_0^3 r_1
- 6 r_0^2 r_1 \, ,
\\[0.4em]
\mathcal{Q}_E(\chi,r_0,r_1)
&=
-16 \chi^5
- \chi^3 \left(14 r_0^3 - 22 r_0^2 - 16 r_0 \right)
\nonumber\\
&\quad
- \chi \left(2 r_0^5 - 14 r_0^4 + 24 r_0^3 - 4 r_0^2 - 8 r_0 r_1 \right) \, .
\end{align}
\end{subequations}

For the first-order tidal correction to the angular momentum, we find
\begin{equation}
\label{eq:Lisco_perturbed}
L^\sigma_{1\,{\rm ISCO}}
=
M \, \frac{
\mathcal{P}_L(\chi,r_0,r_1)
+ \sigma\, \sqrt{r_0}\, \mathcal{Q}_L(\chi,r_0,r_1)
}{
2\, r_0^{11/4}
\left[\sqrt{r_0}(r_0 - 3) + 2 \chi \sigma \right]^{3/2}
} \, ,
\end{equation}
where the polynomials $\mathcal{P}_L$ and $\mathcal{Q}_L$ read
\begin{subequations}
\label{eqs:PQ_Lisco_perturbed}
\begin{align}
\mathcal{P}_L(\chi,r_0,r_1)
&=
6 \chi^7
+ 6 \chi^5 r_0 (1 - 2 r_0)
- \chi^3 r_0 \left(6 r_0^4 - 14 r_0^3 - 18 r_0^2 + 16 r_0 + 3 r_1 \right)
\nonumber\\
&\quad
- \chi r_0^2 \left(6 r_0^5 - 36 r_0^4 + 44 r_0^3 - 4 r_0^2 - 9 r_0 r_1 + 6 r_1 \right) \, ,
\\[0.4em]
\mathcal{Q}_L(\chi,r_0,r_1)
&=
2 \chi^6 (3 r_0 - 8)
- 2 \chi^4 r_0 \left(3 r_0^2 - 8 \right)
\nonumber\\
&\quad
- \chi^2 r_0 \left(2 r_0^5 + 6 r_0^4 - 22 r_0^3 + 24 r_0^2 + 3 r_0 r_1 - 4 r_0 - 8 r_1 \right)
\nonumber\\
&\quad
- 2 r_0^8
+ 10 r_0^7
- 18 r_0^6
+ 4 r_0^5
+ 12 r_0^4
+ r_0^3 (r_0 - 6) r_1 \, .
\end{align}
\end{subequations}

For the first-order tidal correction to the orbital angular frequency at the ISCO,
we obtain
\begin{equation}
\label{eq:frequency_ISCO_perturbed}
\Omega^{\sigma}_{1\,{\rm ISCO}}
\, M=
\frac{
\mathcal{P}_{\Omega}(\chi,r_0)
+ \sigma\, \mathcal{Q}_{\Omega}(\chi,r_0,r_1)
}{
2 \sqrt{r_0}\, \left(r_0^{3/2} + \sigma \chi \right)^2
} \, ,
\end{equation}
where the functions $\mathcal{P}_{\Omega}$ and $\mathcal{Q}_{\Omega}$ are given by
\begin{subequations}
\label{eqs:PQ_frequencyisco_perturbed}
\begin{align}
\mathcal{P}_{\Omega}(\chi,r_0)
&=
4 \chi \sqrt{r_0}\,\left(r_0^3 - \chi^2 \right) \, ,
\\[0.4em]
\mathcal{Q}_{\Omega}(\chi,r_0,r_1)
&=
6 \chi^4
- 2 \chi^2 r_0 (r_0 + 3)
- 2 r_0^5
+ 4 r_0^2
- 3 r_0 r_1 \, .
\end{align}
\end{subequations}

Finally, the first-order tidal correction to the redshift invariant reads
\begin{equation}
\label{eq:redshift_ISCO_perturbed}
U^\sigma_{1\,{\rm ISCO}}
=
\frac{
\mathcal{P}_{U}(\chi,r_0,r_1)
+ 2 \sigma \sqrt{r_0}\, \mathcal{Q}_{U}(\chi,r_0,r_1)}{2 r_0^{11/4}
\left[\sqrt{r_0}(r_0 - 3) + 2 \sigma \chi \right]^{3/2}
} \, ,
\end{equation}
where
\begin{subequations}
\label{eqs:PQ_redshiftisco_perturbed}
\begin{align}
\mathcal{P}_{U}(\chi,r_0,r_1)
&=
6 \chi^6
+ 6 \chi^4 r_0 (2 r_0 - 1)
+ \chi^2 r_0 \left(4 r_0^4 - 24 r_0^3 - 6 r_0^2 + 12 r_0 - 3 r_1 \right)
\nonumber\\
&\quad
- r_0^3 \left(6 r_0^3 - 8 r_0^2 + 3 r_1 \right) \, ,
\\[0.4em]
\mathcal{Q}_{U}(\chi,r_0,r_1)
&=
- 6 \chi^5
+ \chi^3 r_0 \left(7 r_0^2 - 9 r_0 + 6 \right)
+ \chi r_0 \left(3 r_0^4 - 3 r_0^3 + 8 r_0^2 - 6 r_0 + 3 r_1 \right) \, .
\end{align}
\end{subequations}
These expressions reveal a strong dependence of the tidal ISCO shifts on both
the black-hole spin and the relative orientation of the orbit. In Fig.~\ref{fig:ISCO} we illustrate the dependence of these quantities on the spin
of the Kerr black hole. We comment further on these results below.
In the non-spinning limit $\chi=0$, all tidal corrections smoothly reduce to those
computed for a tidally deformed Schwarzschild black hole in Ref.~\cite{Camilloni:2023rra}.


\subsection{LR shifts}
\label{subsec:LR_shifts}

Similarly, one can analyze how the tidal environment affects the location and
properties of the LR of a Kerr black hole by solving the following
conditions in terms of the secular Hamiltonian:
\begin{equation}
\label{eq:LR_cond}
\langle H\rangle \big|_{r=r_{\rm LR}} = 0 \, , \qquad
\frac{d\langle H\rangle}{dr}\bigg|_{r=r_{\rm LR}} = 0 \, .
\end{equation}
The solution of these equations yields the LR radius, the impact
parameter, and the orbital frequency, which completely characterize the LR. We
define the impact parameter as
\[
b \equiv \sigma\,\frac{L}{E},
\]
where $\sigma=\pm1$ labels prograde and retrograde orbits, respectively. With this convention, the conserved angular momentum $L$ is positive (negative) for prograde (retrograde) motion, so that $b$ is positive for both orientations and
retains its standard geometric interpretation as a distance.

Up to first order in the perturbative parameter $\eta$, we write
\begin{equation}
\label{eq:LR_corrections_rb}
\begin{aligned}
r^{\sigma}_{\rm LR} &\simeq
r^{\sigma}_{0\,{\rm LR}} + \eta\, r^{\sigma}_{1\,{\rm LR}} \, , \\
b^{\sigma}_{\rm LR} &\simeq
b^{\sigma}_{0\,{\rm LR}} + \eta\, b^{\sigma}_{1\,{\rm LR}} \, , \\
\Omega^{\sigma}_{\rm LR} &\simeq
\Omega^{\sigma}_{0\,{\rm LR}} + \eta\, \Omega^{\sigma}_{1\,{\rm LR}} \, .
\end{aligned}
\end{equation}

By solving Eqs.~\eqref{eq:LR_cond} to zeroth order in the tidal parameter $\eta$, we recover the known results for the radial coordinate, impact parameter, and frequency of the unperturbed LRs of a Kerr black hole~\cite{Chandrasekhar:1985kt, Bardeen:1972fi}
\begin{equation}
    \label{eq:rLR_bLR_unperturbed}
    \frac{r^{\sigma}_{0\,{\rm{LR}}}}{M}  =  2 \left(1+ \cos \left[\frac{2}{3} \arccos (-\sigma \chi)\right]\right) \,,~~
    \frac{b^{\sigma}_{0\, {\rm{LR}}}}{M} = \frac{\hat{r}_{0}^2-2 \sigma \chi \hat{r}_{0}^{1/2} + \chi^2}{\hat{r}_{0}^{3/2}-2\hat{r}_{0}^{1/2} + \sigma \chi} \,,~~
    \Omega^\sigma_{0 \,{\rm{LR}}}  \, M =   \frac{\sigma}{\hat{r}^{3/2}_{0}+ \sigma \chi}\,
\end{equation}
where we recall that $\chi\equiv a/M$. Moreover, to simplify the notation, we defined the dimensionless unperturbed LR radius $\hat{r}_0 \equiv r^{\sigma}_{0\, {\rm{LR}}}/M$. Here, the LR angular frequency is defined analogously as the secularly
averaged quantity $\Omega \equiv \langle u^\psi/u^v \rangle$, evaluated on circular equatorial null orbits.

By solving Eqs.~\eqref{eq:LR_cond} to first order in the tidal parameter $\eta$, we obtain the leading tidal corrections to the LR quantities.

For the first-order shift of the LR radius, we find
\begin{equation}
\label{eq:rLR_perturbed}
r^\sigma_{1\,{\rm LR}}
= M
\frac{
2 \left[
\mathcal{F}(\chi,\hat r_0)
+ \sigma \sqrt{\hat r_0}\, \mathcal{G}(\chi,\hat r_0)
\right]
}{
\hat r_0 \left(
\hat r_0^2 - 6 \hat r_0 - 3 \chi^2 + 8 \sigma \chi \sqrt{\hat r_0}
\right)
} \, ,
\end{equation}
where the functions $\mathcal{F}$ and $\mathcal{G}$ are given by
\begin{subequations}
\label{eqs:FG_rLR_perturbed}
\begin{align}
\mathcal{F}(\chi,\hat r_0)
&=
-3 \chi^6
+ \chi^4(2 \hat r_0^2 - 3 \hat r_0)
+ \chi^2(4 \hat r_0^5 - 16 \hat r_0^4 + 11 \hat r_0^3 + 8 \hat r_0^2)
\nonumber\\
&\quad
+ \left(2 \hat r_0^7 - 11 \hat r_0^6 + 22 \hat r_0^5 - 16 \hat r_0^4 \right) \, ,
\\[0.4em]
\mathcal{G}(\chi,\hat r_0)
&=
8 \chi^5
+ \chi^3(7 \hat r_0^3 - 11 \hat r_0^2 - 8 \hat r_0)
+ \chi\left(\hat r_0^5 - 7 \hat r_0^4 + 12 \hat r_0^3 - 2 \hat r_0^2 \right) \, .
\end{align}
\end{subequations}

For the first-order tidal correction to the impact parameter, we obtain
\begin{equation}
\label{eq:bLR_perturbed}
b^\sigma_{1\,{\rm LR}}
= M
\frac{
\mathcal{R}_b(\chi,\hat r_0)
+ \sigma \sqrt{\hat r_0}\, \mathcal{K}_b(\chi,\hat r_0)
}{
(\hat r_0 - 2)\sqrt{\hat r_0} + \sigma \chi
} \, ,
\end{equation}
where the functions $\mathcal{R}_b$ and $\mathcal{K}_b$ read
\begin{subequations}
\label{eqs:RK_bLR_perturbed}
\begin{align}
\mathcal{R}_b(\chi,\hat r_0)
&=
2 \chi^4
+ \chi^2(3 \hat r_0^3 - 5 \hat r_0^2)
+ \left(\hat r_0^5 - 3 \hat r_0^4 + 4 \hat r_0^3 - 2 \hat r_0^2 \right) \, ,
\\[0.4em]
\mathcal{K}_b(\chi,\hat r_0)
&=
2 \chi^3(\hat r_0 - 2)
- 2 \chi\, \hat r_0 (\hat r_0^2 - 2) \, .
\end{align}
\end{subequations}
Finally, we find that the angular frequency of the LR receives a tidal
correction given by
\begin{equation}
\label{eq:frequency_LR_perturbed}
\Omega^{\sigma}_{1\,{\rm LR}} \, M
=
\frac{
\mathcal{R}_{\Omega}(\chi,\hat r_0,\hat r_1)
+ \sigma\, \sqrt{\hat r_0}\,
\mathcal{K}_{\Omega}(\chi,\hat r_0,\hat r_1)
}{
(\hat r_0^2 - 2 \hat r_0 + \chi^2 )\,
\left(\hat r_0^{3/2} + \sigma \chi \right)^2
} \, ,
\end{equation}
where $\hat r_1 \equiv r^\sigma_{1\,{\rm LR}}/M$, and the functions
$\mathcal{R}_{\Omega}$ and $\mathcal{K}_{\Omega}$ read
\begin{subequations}
\label{eqs:RK_frequencyLR_perturbed}
\begin{align}
\mathcal{R}_{\Omega}(\chi,\hat r_0,\hat r_1)
&=
6 \chi^5
+ \chi^3\left(9 \hat r_0^{3} - 13 \hat r_0^{2} - 4 \hat r_0 \right)
\nonumber\\
&\quad
+ \chi\left(3 \hat r_0^{5} - 11 \hat r_0^{4}
+ 12 \hat r_0^{3} - 2 \hat r_0^{2}
- 4 \hat r_0 \hat r_1 \right) \, ,
\\[0.4em]
\mathcal{K}_{\Omega}(\chi,\hat r_0,\hat r_1)
&=
4 \chi^4(\hat r_0 - 3)
+ \chi^2\left(3 \hat r_0^{4}
- 17 \hat r_0^{3}
+ 10 \hat r_0^{2}
+ 16 \hat r_0 \right)
\nonumber\\
&\quad
+ \hat r_0^{6}
- 9 \hat r_0^{5}
+ 22 \hat r_0^{4}
- 14 \hat r_0^{3}
- 4 \hat r_0^{2}
- 2 \hat r_0^{2} \hat r_1
+ 6 \hat r_0 \hat r_1 \, .
\end{align}
\end{subequations}
Note that, in the non-spinning limit $\chi=0$, this expression correctly reduces to the first-order tidal correction to the light-ring frequency of a Schwarzschild black hole derived in Ref.~\cite{Camilloni:2023rra}.

The analytical results obtained for the tidal deformations of the ISCO and the LR are derived for a generic external tidal environment. The only assumptions are that $M  \ll \mathcal{R}$, and that the tidal perturbation is retained only up to the leading quadrupolar order.

\begin{figure}[ht]
    \label{fig:ISCO}
    \centering
    \includegraphics[width=1.0\linewidth]{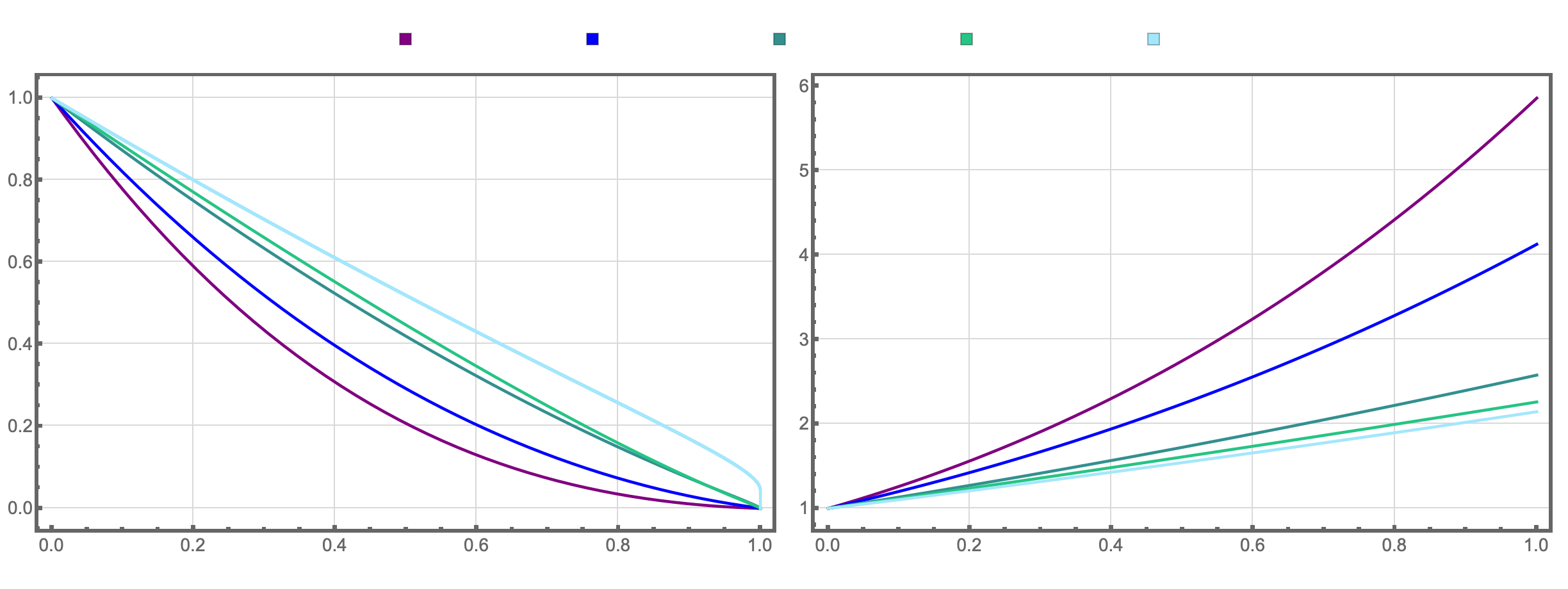}
    \put(-345,-1){\scriptsize $\chi$}
    \put(-120,-1){\scriptsize $\chi$}
    \put(-330,157){\scriptsize {$\rm{r}_1$/$\rm{r}_1^{\rm S}$}}
    \put(-275,157){\scriptsize {$\rm{L}_1$/$\rm{L}_1^{\rm S}$}}
    \put(-220,157){\scriptsize {$\rm{E}_1$/$\rm{E}_1^{\rm S}$}}
    \put(-167,157){\scriptsize {$\rm{\Omega}_{1}/{\rm{\Omega}_{1}^{\rm S}}$}}
    \put(-113,157){\scriptsize {$\rm{U}_{1}$/$\rm{U}_{1}^{\rm S}$}}
    \caption{Secular tidal perturbations for the radius $r$, energy $E$, angular momentum $L$, frequency $\Omega$, and redshift factor $U$ of a massive test particle moving on the ISCO of a tidally deformed Kerr black hole. The shifts are shown as functions of the dimensionless spin $\chi$ and are normalized to the Schwarzschild ($\chi=0$) case. The left panel corresponds to prograde motion, while the right panel corresponds to retrograde motion.
}
\end{figure}

\begin{figure}[ht]
    \label{fig:LR}
    \centering
    \includegraphics[width=1.0\linewidth]{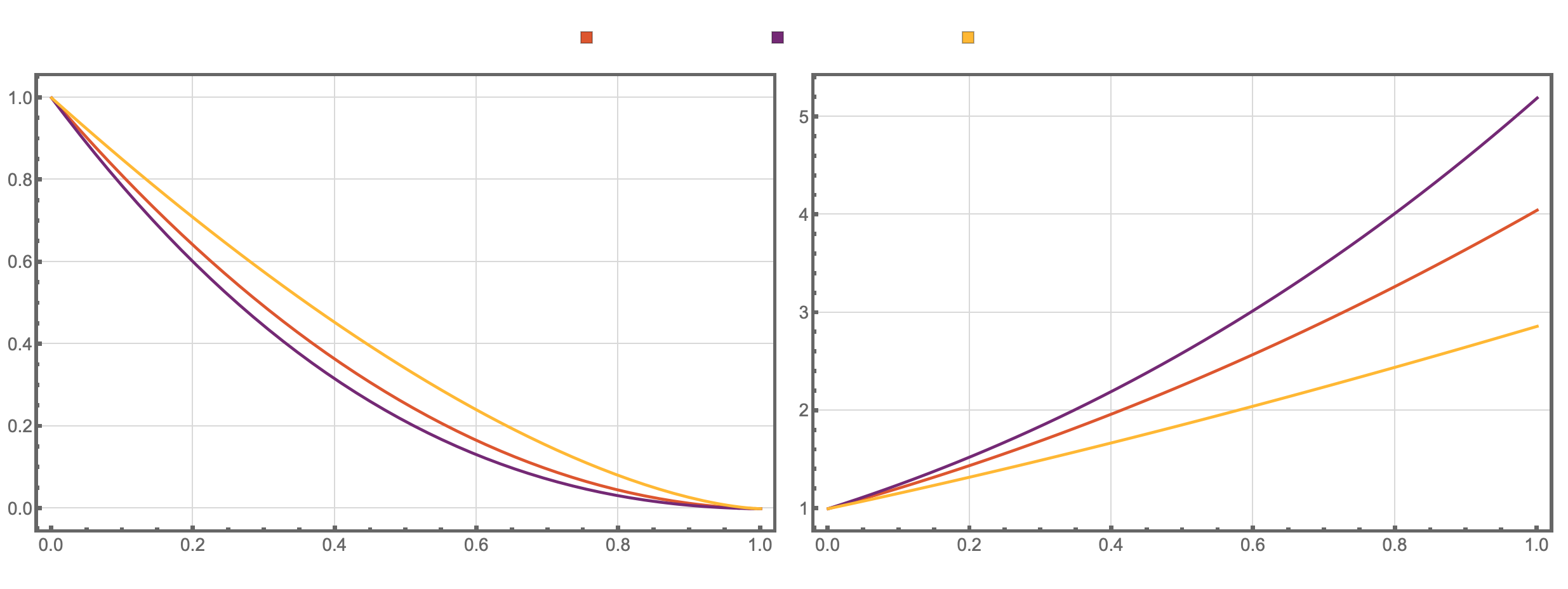}
    \put(-345,0){\scriptsize $\chi$}
    \put(-120,0){\scriptsize $\chi$}
    \put(-276,158){\scriptsize {$\rm{r}_1$/$\rm{r}_1^{\rm S}$}}
    \put(-221,158){\scriptsize {$\rm{b}_1$/$\rm{b}_1^{\rm S}$}}
    \put(-166,158){\scriptsize {$\rm{\Omega}_1$/$\rm{\Omega}_1^{\rm S}$}}
    \caption{ Secular tidal perturbations of the radius $r$, impact parameter $b$, and frequency $\Omega$ of a massless test particle moving on the LR of a tidally deformed Kerr black hole. The shifts are shown as functions of the dimensionless spin $\chi$ and are normalized to the Schwarzschild ($\chi=0$) case. The left panel corresponds to prograde motion, while the right panel corresponds to retrograde motion.}
\end{figure}

We illustrate these results in Figs.~\ref{fig:ISCO} and~\ref{fig:LR}, where we show the behavior of the secular tidal corrections, normalized to the Schwarzschild result ($\chi = 0$) of Ref.~\cite{Camilloni:2023rra}, as functions of the dimensionless spin parameter $\chi$. In both cases, the plots reveal a pronounced dependence of these tidal corrections on the black-hole spin.

Specifically, the left panels of both figures show that, for prograde orbits,
tidal shifts are increasingly suppressed as the black-hole spin grows, relative
to the corresponding non-spinning case. In contrast, the right panels show that for retrograde orbits the trend is reversed: increasing the black-hole spin
leads to progressively enhanced tidal deformations.

This behavior is consistent across all orbital quantities considered for both the ISCO and the LR, and admits a clear physical interpretation.
As originally discussed by Bardeen and collaborators~\cite{Bardeen:1972fi},
increasing the black-hole spin causes prograde circular orbits to move closer to
the event horizon $r_+$, while retrograde orbits are pushed to larger radii.
As a consequence, prograde ISCO and LR orbits probe regions of stronger
gravitational binding, where the coupling to a fixed external tidal field is
effectively reduced, leading to a suppression of tidal effects as the spin increases. A qualitatively similar suppression of tidal effects in the extremal limit was observed in Ref.~\cite{Grilli:2024fds} for charged, non-rotating black-hole binaries.
Conversely, retrograde orbits reside at larger radii for higher spins, where the tidal interaction is stronger, resulting in enhanced tidal deformations.

\begin{figure}[ht]
\label{fig:absolute_radius_shifts}
    \centering
    \includegraphics[width=1.0\linewidth]{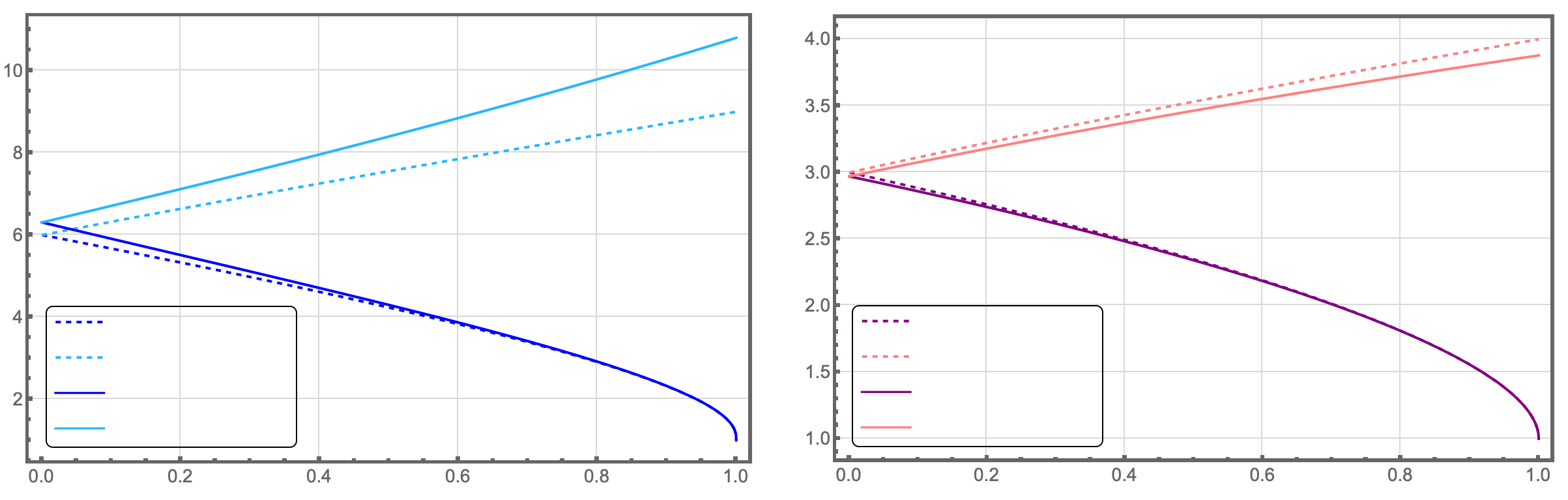}
    \put(-348,-5){\scriptsize $\chi$}
    \put(-115,-5){\scriptsize $\chi$}
    \put(-460,66){\rotatebox{90} {\scriptsize $r/M$}}
    \put(-231,66){\rotatebox{90} {\scriptsize $r/M$}}
     \put(-420,50){\tiny $r_{0}$ ($\sigma$=+1)}
    \put(-420,40){\tiny $r_{0}$ ($\sigma$=--1)}
    \put(-420,30){\tiny $r_{0}$+$\eta$$r_1$($\sigma$=+1)}
    \put(-420,20){\tiny $r_{0}$+$\eta$$r_1$($\sigma$=--1)}
     \put(-186,50){\tiny $r_{0}$ ($\sigma$=+1)}
    \put(-186,40){\tiny $r_{0}$ ($\sigma$=--1)}
    \put(-186,30){\tiny $r_{0}$+$\eta$$r_1$($\sigma$=+1)}
    \put(-186,20){\tiny $r_{0}$+$\eta$$r_1$($\sigma$=--1)}
 \caption{ Secular tidal shift of the radial coordinate for ISCO (left panel) and LR (right panel), as functions of the dimensionless black-hole spin $\chi$. Dashed lines denote the unperturbed Kerr values, while solid lines denote the tidally corrected radii $r_0^\sigma+\eta\, r_1^\sigma$. $\sigma= + 1$ denotes prograde orbits while $\sigma=-1$ refers to retrograde orbits. We use $\eta=10^{-4}$ for the ISCO and $\eta=10^{-3}$ for the LR for visualization purposes.}
\end{figure}

To further describe the physical impact of the tidal perturbation, in Fig.~\ref{fig:absolute_radius_shifts} we plot the perturbed radii
$r^{\sigma}_{0\,{\rm ISCO}}+\eta \, r^{\sigma}_{1\,{\rm ISCO}}$ and
$r^{\sigma}_{0\,{\rm LR}}+\eta\,  r^{\sigma}_{1\,{\rm LR}}$
as a function of the dimensionless black-hole spin $\chi$, and compare them with the corresponding unperturbed Kerr values $r^{\sigma}_{0\,{\rm ISCO}}$ and
$r^{\sigma}_{0\,{\rm LR}}$. For visualization purposes we choose $\eta=10^{-4}$ for the ISCO and
$\eta=10^{-3}$ for the LR. These values are chosen to make the shifts visible in the plot; physically realistic values are typically smaller (see App.~\ref{app:eta_estimate}).

From Fig.~\ref{fig:absolute_radius_shifts} one can clearly see that, for increasing values of the black-hole spin, the tidal perturbation drives the ISCO and LR radii for prograde orbits closer to their unperturbed values. For retrograde orbits, the behavior is reversed, and the difference between the perturbed and unperturbed radii increases as one approaches the extremal limit $\chi \to 1$. This is consistent with the discussion above: increasing spin pushes prograde (retrograde) orbits to smaller (larger) radii, where tidal effects are correspondingly suppressed (enhanced). The same trend applies to all orbital quantities considered, including the frequency and redshift shifts. We provide plots for the tidal corrections of the remaining observables relevant for the ISCO and LR orbits (energy, angular momentum, impact parameter, frequency, and redshift) in App.~\ref{app:absolute_tidal_shifts}. 

Finally, while several analyses in the literature restrict attention to the
subextremal regime $0<|\chi|<1$, we find no indication of pathological behavior
as $|\chi|\to 1$ in the static ($\omega=0$) sector considered here. In particular, in the prograde case, the unperturbed ISCO and LR radii approach the horizon in the extremal limit, $r_{+}|_{\chi=1}=M$, and the corresponding first-order tidal shifts vanish. Thus, the tidally corrected quantities coincide smoothly with their unperturbed extremal Kerr values. In the retrograde case, the orbits remain at finite radius outside the horizon in the extremal limit, and the corresponding tidal shifts approach finite, nonzero values. We have explicitly evaluated these limits, finding regular behavior for all orbital quantities, although the magnitude of the shifts is enhanced relative to the prograde case.

We note, however, that to make a more quantitative comment about geodesic motion in the extremal case, one needs to perform a more extensive study of the near-horizon region~\cite{Kapec:2019hro}. Our analysis should therefore be understood within the present far-region, static, quadrupolar framework. Within these assumptions, the metric perturbation reconstructed in the ORG and the associated quantities derived from the secular Hamiltonian behave smoothly in the extremal limit.
%
 

\section{Conclusions}
\label{Sec: Conclusions}

In this paper, we developed an analytic framework to describe the tidal deformation of rotating black holes and applied it to the study of the secular dynamics of test particles in the SF regime. By solving the Teukolsky master equation in the Hartle--Hawking tetrad for static modes and employing metric reconstruction techniques in the outgoing radiation gauge, we derived fully explicit expressions for the quadrupolar metric perturbation of a Kerr black hole subject to slowly varying external tidal fields. In the
non-spinning limit, our results smoothly reduce to the known expressions for
tidally deformed Schwarzschild black holes~\cite{Binnington:2009bb}. All results
are obtained in the near-zone regime $M \ll \mathcal{R}$ and at leading quadrupolar order in the tidal deformation.

Using the reconstructed metric, we investigated the secular dynamics of test
particles in the tidally deformed Kerr spacetime and derived the corresponding secular Hamiltonian to first order in the tidal perturbation. This allowed us to compute tidal-induced shifts in the location and properties of the innermost
stable circular orbit and the light ring. We found that these shifts exhibit a strong dependence on the black-hole spin and on the relative orientation between the orbital angular momentum and the spin. In particular, tidal effects are
suppressed for prograde orbits around rapidly rotating black holes, while they are significantly enhanced for retrograde orbits. These results highlight the role of spin-tidal couplings in shaping the SF orbital dynamics of
matter and radiation in the near-horizon region, and extend to spinning black holes the SF tidal effects previously identified in charged black-hole spacetimes~\cite{Grilli:2024fds}.

Our results represent a step toward a more complete understanding of the
relativistic tidal response of spinning black holes. In particular, the fully analytic nature of our framework provides controlled access to spin--tidal couplings and secular SF effects in rotating black-hole spacetimes, bridging perturbative tidal analyses and applications relevant to gravitational-wave observations.

Several directions for future work emerge from this analysis. An important
extension is the inclusion of higher-order tidal multipoles, as well as the study of time-dependent tidal perturbations. It would also be relevant to
compute the associated energy and angular-momentum fluxes, following existing
approaches in black-hole perturbation theory~\cite{Poisson:2004cw,
Chatziioannou:2012gq, Berens:2024czo}. These extensions are likely to be
particularly relevant for the modeling of EMRIs and for future tests of general
relativity with space-based gravitational-wave detectors such as LISA.

A further extension is to move beyond the circular and equatorial approximation adopted in this paper. The present framework can be extended to generic bound geodesics in a tidally deformed Kerr spacetime, including eccentric and inclined
orbits, by employing the general secular-averaging procedure given in Eq.~\eqref{eq:secularaverage_general}. This would allow one to investigate how
external tidal fields influence orbital precession, resonances, and long-term
secular evolution in EMRI systems beyond the circular limit.

A related but distinct application concerns hierarchical triple systems, in
which the external tidal field is generated by a third massive black hole.
Recent work has shown that tidal interactions in such systems can significantly
affect the eccentricity evolution and merger timescale of compact binaries
treated within post-Newtonian theory~\cite{Camilloni:2023xvf,
Cocco:2025adu, Cocco:2025udb}. The tidally deformed Kerr metric derived here
provides a SF framework to study analogous effects when the inner object is modeled as a test particle orbiting a spinning black hole, thereby enabling a systematic comparison between EMRI and comparable-mass regimes.

Finally, from a more fundamental perspective, it is desirable to understand the physics of near-extremal and extremal black holes. These have been shown to be very sensitive to perturbations, which can be amplified drastically close to the event horizon \cite{Horowitz:2023xyl}. It would be interesting to explore if, in the extremal limit, there exist observables where tidal perturbations are similarly amplified, even within general relativity.
We plan to explore these extensions in future work.


\section*{Acknowledgements}

M.~Cocco, G.~Grignani, and M.~Orselli acknowledge financial support from the
Italian Ministry of University and Research (MUR) through the program
``Dipartimenti di Eccellenza 2018--2022'' (Grant SUPER-C).
G.~Grignani and M.~Orselli also acknowledge support from the
``Fondo di Ricerca d'Ateneo'' 2023 (GraMB) of the University of Perugia, and from the Italian Ministry of University and Research (MUR) via PRIN~2022ZHYFA2,
\emph{GRavitational wavEform models for coalescing compAct binaries with
eccenTricity} (GREAT). M.~Cocco, T.~Harmark, M.~van~de~Meent, and M.~Orselli acknowledge support from the ``Center of Gravity'', a Center of Excellence funded by the Danish National
Research Foundation under Grant No.~DNRF184.
D.~Pereñiguez is further supported by NSF
Grants No. AST-2307146, PHY-2513337, PHY-090003,
and PHY-20043, by NASA Grant No. 21-ATP21-0010,
by John Templeton Foundation Grant No. 62840, by
the Simons Foundation, and by Italian Ministry of Foreign Affairs and International Cooperation Grant No.
PGR01167.
M.~van~de~Meent further acknowledges financial support from the VILLUM Foundation
(Grant No.~VIL37766), the DNRF Chair program (Grant No.~DNRF162) funded by the
Danish National Research Foundation, and the European Union’s Horizon ERC Synergy
Grant \emph{Making Sense of the Unexpected in the Gravitational-Wave Sky}
(GWSky--101167314).


\appendix
\section{Order-of-magnitude estimate of the tidal parameter $\eta$}
\label{app:eta_estimate}

In this appendix, we provide an order-of-magnitude estimate of the tidal parameter $\eta$ defined in Eq.~\eqref{eq:eta}, to assess both the regime of validity of our perturbative treatment and the expected magnitude of the tidal corrections discussed in Sec.~\ref{sec:ISCO_LR_shifts}.

\subsection{Analytical estimates}

We model the tidal perturber as a Kerr black hole of mass $M_\star$ and spin $a_\star$, providing a realistic realization of a hierarchical triple system. 
The tidal parameter $\eta$, defined in Eq.~\eqref{eq:eta}, is given by
\begin{equation}
\eta \equiv -\frac{M^2}{2}\langle \mathcal{E}^{\rm q}\rangle \, .
\end{equation}
Assuming $M_\star \gg M$, one finds~\cite{Camilloni:2023rra}
\begin{equation}
\label{eq:eta_SF_app}
\eta = \frac{M_\star M^2}{16 r_\star^3} \left(1+ 3\frac{K_\star}{r_\star^2}\right),
\end{equation}
where $K_\star$ is the Carter constant, and $r_\star$ is the Boyer--Lindquist radial coordinate in the Kerr spacetime of the tidal perturber.

In the weak-field (WF) regime $r_\star \gg M_\star$, Eq.~\eqref{eq:eta_SF_app} reduces to
\begin{equation}
\label{eq:eta_WF_app}
    \eta_{\rm WF} = \frac{M^2}{4}\frac{M_\star}{r_\star^3}\,.
\end{equation}

In the strong-field (SF) regime $r_\star \sim M_\star$, relevant for relativistic hierarchical triples, we further specialize to circular equatorial motion at the ISCO of the tidal Kerr spacetime $(M_\star,a_\star)$. In this case,
\begin{equation}
\label{eq:eta_SF_ISCO_app}
\eta_{\rm SF,ISCO} = \frac{M^2}{2}\frac{M_\star}{( r^\sigma_{\star \, \rm ISCO})^3} = \left(\frac{M}{M_\star}\right)^2 \frac{1}{2\,f(a_\star/M_\star)} \,,
\end{equation}
where
\[ 
f(a_\star/M_\star) = \left[ 3+Z_2-\sigma\sqrt{(3-Z_1)(3+Z_1+2Z_2)} \right]^3 ,
\]
with
\[ 
Z_1 = 1+\left(1-\left(\frac{a_\star}{M_\star}\right)^2\right)^{1/3}
\left[\left(1+\frac{a_\star}{M_\star}\right)^{1/3}
+\left(1-\frac{a_\star}{M_\star}\right)^{1/3}\right],
\qquad Z_2 = \sqrt{Z_1^2+3\left(\frac{a_\star}{M_\star}\right)^2}\,,
\]
and $\sigma=\pm1$ for prograde and retrograde ISCOs, respectively.

These expressions show that $\eta$ remains small either in the weak-field regime, due to large separations, or in the SF regime, where it is suppressed by the mass hierarchy $ M/M_\star \ll 1$.

\subsection{Numerical estimates}

We now provide numerical estimates of $\eta$ based on the analytical expressions in Eqs.~\eqref{eq:eta_WF_app} and~\eqref{eq:eta_SF_ISCO_app}.

\begin{figure}[ht]
    \centering
    \includegraphics[width=1.0\linewidth]{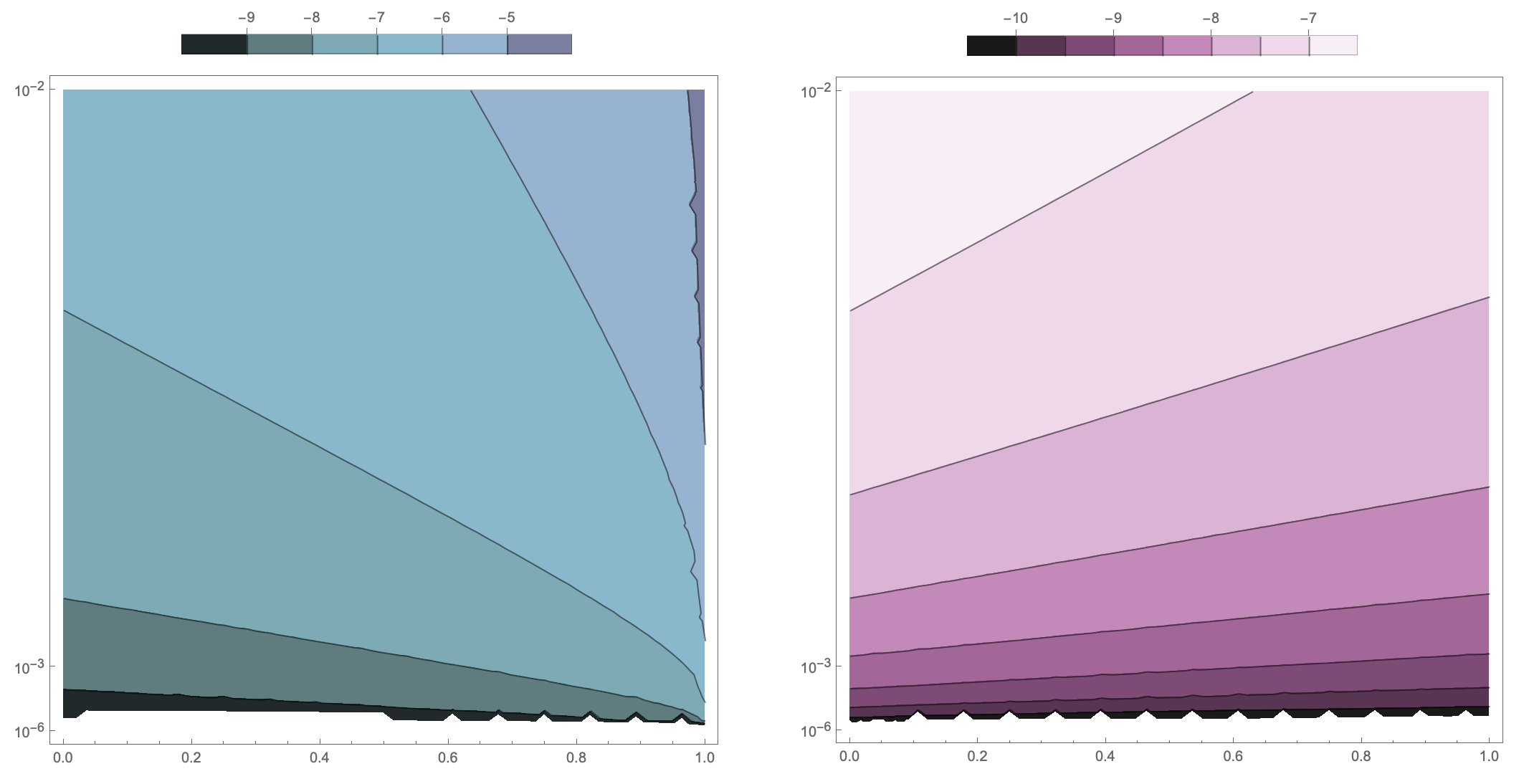}
    \put(-355,-6){\scriptsize $a_\star/M_\star$}
    \put(-120,-6){\scriptsize $a_\star/M_\star$}
    \put(-450,95){\rotatebox{90} {\scriptsize $M/M_\star$}}
    \put(-220,95){\rotatebox{90} {\scriptsize $M/M_\star$}}
\caption{ Contour plots of $\log_{10}\eta_{\rm SF,ISCO}$ from Eq.~\eqref{eq:eta_SF_ISCO_app}, as a function of the spin $a_\star/M_\star$ of the Kerr perturber and the mass ratio $M/M_\star$, for a system located at the ISCO of the tidal Kerr spacetime $(M_\star,a_\star)$. 
Left: prograde orbits ($\sigma=+1$). Right: retrograde orbits ($\sigma=-1$).
}
\label{fig:eta_SF_ISCO_contour}
\end{figure}

Fig.~\ref{fig:eta_SF_ISCO_contour} shows contour plots of $\log_{10}\eta_{\rm SF,ISCO}$ computed from Eq.~\eqref{eq:eta_SF_ISCO_app}, as a function of the dimensionless spin $a_\star/M_\star$ of the perturber Kerr black hole and the mass ratio $M/M_\star$. 
In the SF regime, where the EMRI system is assumed to lie at the ISCO of the external Kerr spacetime $(M_\star,a_\star)$, $\eta$ remains perturbatively small ($\eta \ll 1$) throughout the parameter space relevant for hierarchical triples. In particular, for $10^{-6} \lesssim M/M_\star \lesssim 10^{-2}$, one finds typical values $\eta \lesssim 10^{-5}$, with the largest values attained in prograde, near-extremal configurations. This reflects the scaling
$\eta_{\rm SF,ISCO}\propto (r_{\star \, \rm ISCO}^\sigma)^{-3}$ in Eq.~\eqref{eq:eta_SF_ISCO_app}:
prograde motion probes smaller radii in the tidal Kerr spacetime, leading to larger values of the effective tidal parameter $\eta$.

\begin{figure}[ht]
    \centering
    \includegraphics[width=0.7\linewidth]{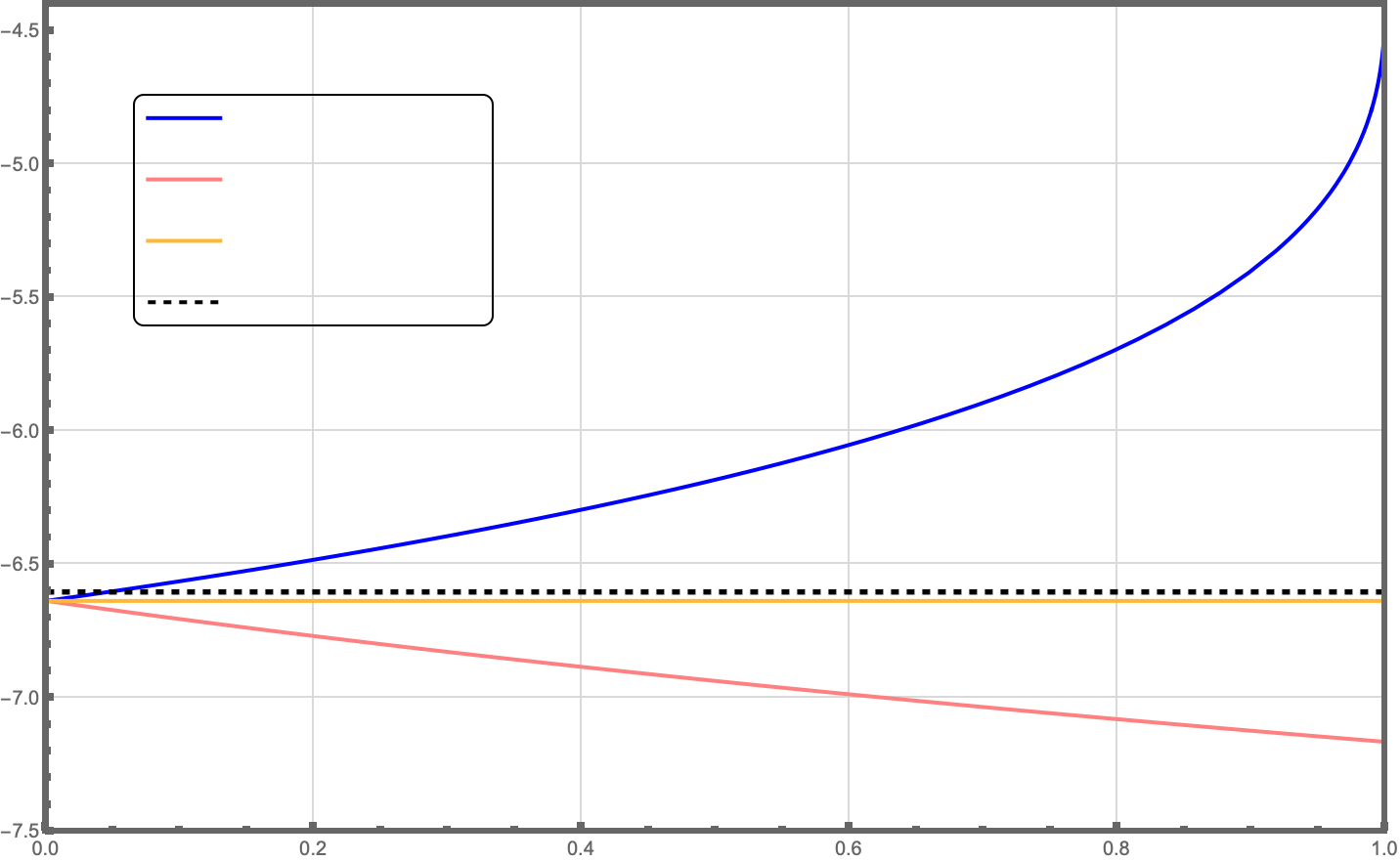}
    \put(-165,-10){\scriptsize $a_\star/M_\star$}
    \put(-330,85){\rotatebox{90} {\scriptsize $\log_{10}\eta$}}
    \put(-264,167){\tiny $\eta_{\rm SF, ISCO ~(\sigma=+1)}$}
    \put(-264,153){\tiny $\eta_{\rm SF, ISCO ~(\sigma=-1)}$}
    \put(-264,139){\tiny $\eta_{\rm SF, ISCO ~(a_\star=0)}$}
    \put(-264,126){\tiny $\eta_{\rm WF}$}
\caption{ Representative estimates of $\log_{10}\eta$ as a function of the spin $a_\star/M_\star$ of the Kerr perturber. 
The solid curves correspond to the strong-field estimate $\eta_{\rm SF,ISCO}$ from Eq.~\eqref{eq:eta_SF_ISCO_app} for prograde ($\sigma=+1$), retrograde ($\sigma=-1$), and Schwarzschild ($a_\star=0$) configurations, evaluated at $M/M_\star=10^{-2}$. 
The dashed curve shows the weak-field estimate $\eta_{\rm WF}$ from Eq.~\eqref{eq:eta_WF_app}, evaluated for $M/M_\star=1$ and $r_\star = 100 M_\star$.
}
\label{fig:etaSF_WF_mu_slices}
\end{figure}

The dependence on the spin $a_\star$ is further illustrated in Fig.~\ref{fig:etaSF_WF_mu_slices}, where we show one-dimensional slices of $\log_{10}\eta$ obtained from Eq.~\eqref{eq:eta_SF_ISCO_app} in the SF case and from Eq.~\eqref{eq:eta_WF_app} for the WF estimate. In the SF regime, $\eta$ increases significantly with the spin for prograde motion, while it decreases for retrograde configurations. The Schwarzschild case ($a_\star=0$) provides an intermediate reference.

For comparison, the WF estimate produces systematically smaller values even when the masses are comparable, highlighting that the smallness of $\eta$ can arise from two distinct physical mechanisms: large separations in the WF regime and mass hierarchy in the SF regime.

In both regimes, we consider parameter values that yield near-maximal values of $\eta$ within their domains of validity. In the SF case, this corresponds to $M/M_\star \sim 10^{-2}$ and placing the system at the ISCO of the tidal Kerr spacetime. In the WF case, we take $M/M_\star \sim 1$ and $r_\star \sim 100 M_\star$.

Overall, these results indicate that a representative value $\eta \sim 10^{-5}$ provides a physically realistic benchmark for the magnitude of tidal effects in the SF regime considered in this work.
%


\section{Tidal shifts of orbital quantities}
\label{app:absolute_tidal_shifts}

In this appendix, we present plots for the tidal corrections of the orbital quantities discussed in Sec.~\ref{sec:ISCO_LR_shifts}.

Specifically, we display the perturbed quantities $X = X_0 + \eta \, X_1$ for all relevant observables: energy, angular momentum, frequency, and redshift (ISCO), as well as the impact parameter and frequency (LR). Unless otherwise specified, retrograde quantities that become negative (i.e., the frequency and angular momentum) are displayed in absolute value.

These plots confirm the qualitative behavior discussed in Sec.~\ref{sec:ISCO_LR_shifts}: the first-order tidal corrections are suppressed for prograde orbits and enhanced for retrograde ones. This behavior is primarily driven by the spin dependence of the orbital radius.

We also highlight some distinctive features of the frequency and redshift, which exhibit nontrivial behavior in the extremal regime.

\begin{figure}[ht]
    \centering
    \includegraphics[width=1.0\linewidth]{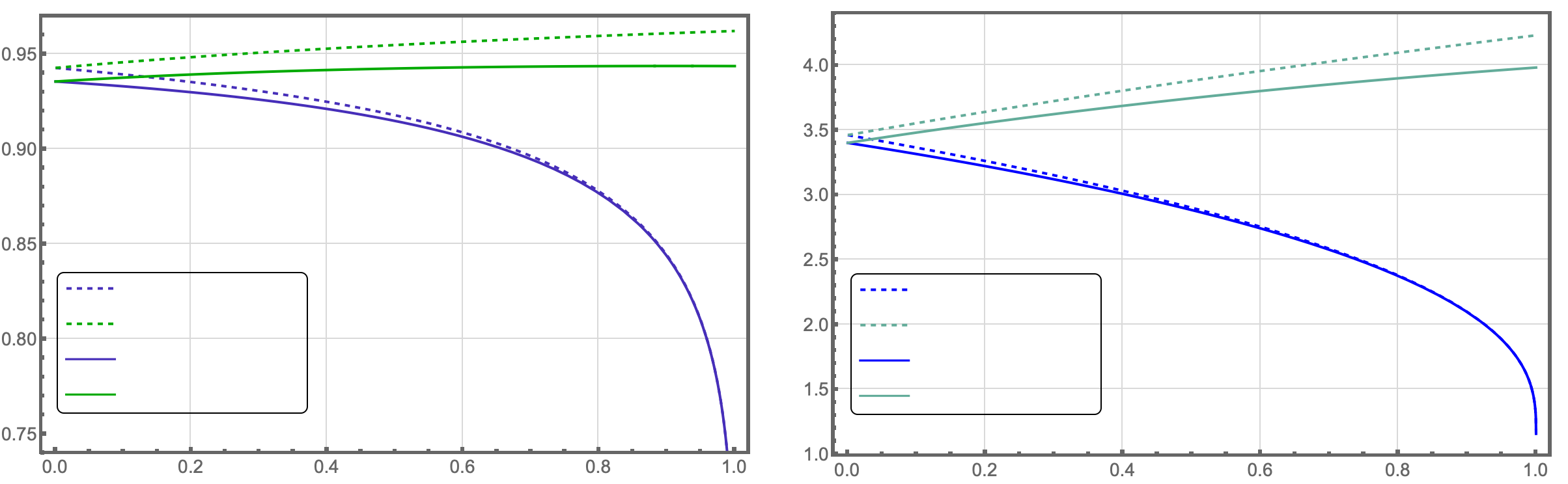}
    \put(-348,-5){\scriptsize $\chi$}
    \put(-115,-5){\scriptsize $\chi$}
    \put(-462,69){\rotatebox{90} {\scriptsize $E$}}
    \put(-231,69){\rotatebox{90} {\scriptsize $L/M$}}
     \put(-418,55){\tiny $\rm E_{0}$ ($\sigma$=+1)}
    \put(-418,45){\tiny $\rm E_{0}$ ($\sigma$=--1)}
    \put(-418,34){\tiny $\rm E_{0}$+$\eta$$\rm E_1$($\sigma$=+1)}
    \put(-418,24){\tiny $\rm E_{0}$+$\eta$$\rm E_1$($\sigma$=--1)}
     \put(-188,55){\tiny $\rm L_{0}$ ($\sigma$=+1)}
    \put(-189,45){\tiny |$\rm L_{0}$| ($\sigma$=--1)}
    \put(-188,34){\tiny $\rm L_{0}$+$\eta$$\rm L_1$($\sigma$=+1)}
    \put(-189.5,24){\tiny |$\rm L_{0}$+$\eta$$\rm L_1$|($\sigma$=--1)}
 \caption{ISCO energy (left panel) and angular momentum (right panel), as functions of the dimensionless black-hole spin $\chi$. Dashed lines denote the unperturbed Kerr values, while solid lines denote the tidally corrected quantities $E_0^\sigma+\eta\, E_1^\sigma$ and $L_0^\sigma+\eta  \, L_1^\sigma$. Retrograde ($\sigma=-1$) orbits display deviations from the unperturbed Kerr values with increasing spin, while prograde ($\sigma=+1$) orbits tend to the unperturbed values as the spin increases. We use $\eta=10^{-4}$ for visualization purposes. 
}
\label{fig:E_and_L_ISCO_tidal_absolute}
\end{figure}

For the ISCO, the behavior of the energy and angular momentum closely follows that of the orbital radius, as shown in Fig.~\ref{fig:E_and_L_ISCO_tidal_absolute}: both quantities tend toward the unperturbed values for prograde orbits as the dimensionless spin $\chi$ increases, whereas for retrograde orbits their deviation from the unperturbed values becomes more pronounced.

\begin{figure}[ht]
    \centering
    \includegraphics[width=1.0\linewidth]{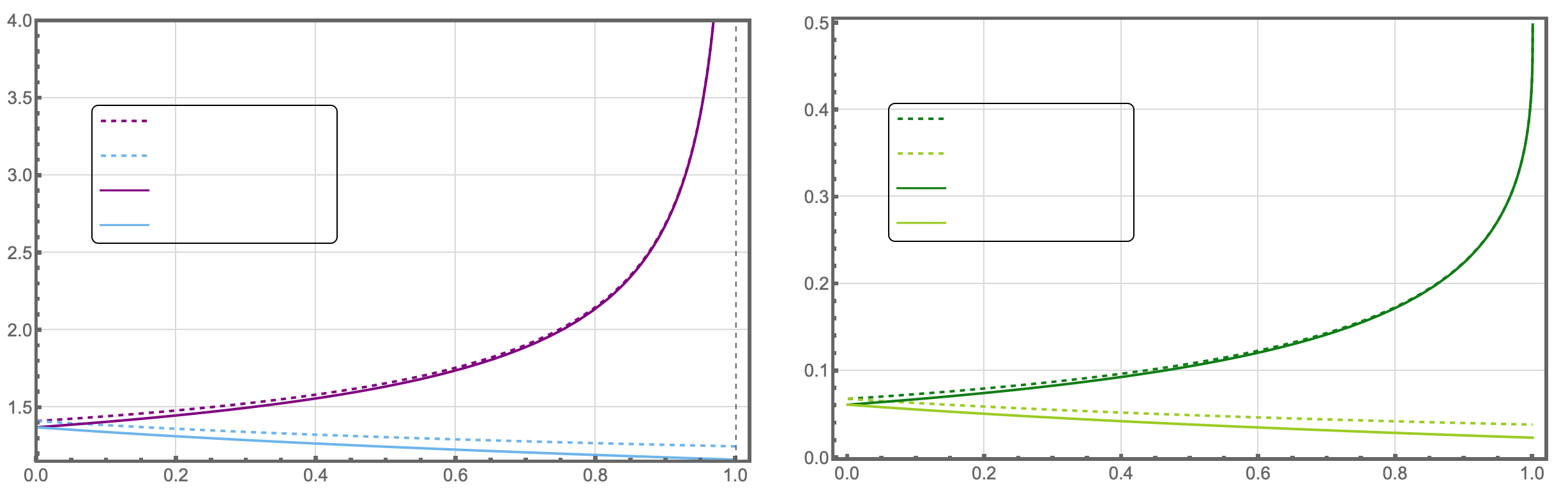}
    \put(-348,-5){\scriptsize $\chi$}
    \put(-115,-5){\scriptsize $\chi$}
    \put(-461,66){\rotatebox{90} {\scriptsize $U$}}
    \put(-230,66){\rotatebox{90} {\scriptsize $\Omega\,M$}}
     \put(-408,105){\tiny $\rm U_{0}$ ($\sigma$=+1)}
    \put(-408,95){\tiny $\rm U_{0}$ ($\sigma$=--1)}
    \put(-408.5,84){\tiny $\rm U_{0}$+$\eta$$\rm U_1$($\sigma$=+1)}
    \put(-408,74){\tiny $\rm U_{0}$+$\eta$$\rm U_1$($\sigma$=--1)}
     \put(-178,105){\tiny $\Omega_{0}$ ($\sigma$=+1)}
    \put(-179,95){\tiny |$\Omega_{0}$| ($\sigma$=--1)}
    \put(-178.5,85){\tiny $\Omega_{0}$+$\eta$$\Omega_1$($\sigma$=+1)}
    \put(-181,75){\tiny |$\Omega_{0}$+$\eta$$\Omega_1$|($\sigma$=--1)}
 \caption{ ISCO redshift (left panel) and frequency (right panel), as functions of the dimensionless black-hole spin $\chi$. Dashed lines denote the unperturbed Kerr values, while solid lines denote the tidally corrected quantities $U_0^\sigma+\eta \, U_1^\sigma$ and $\Omega_0^\sigma+\eta \,\Omega_1^\sigma$. Although the unperturbed Kerr values increase toward the extremal limit for prograde ($\sigma=+1$) orbits, the corresponding tidal corrections become progressively suppressed. For retrograde ($\sigma=-1$) orbits, the opposite behavior is observed. For retrograde motion, we display the absolute value of the frequency. We use $\eta=10^{-4}$ for visualization purposes.
 }
\label{fig:U_and_Omega_ISCO_tidal_absolute}
\end{figure}

In contrast, the frequency and redshift exhibit a qualitatively different behavior, as shown in Fig.~\ref{fig:U_and_Omega_ISCO_tidal_absolute}. In the prograde case, both quantities increase as the extremal limit $\chi \to 1$ is approached, reflecting the behavior of the corresponding unperturbed Kerr quantities. In particular, the frequency approaches the angular velocity of the horizon $\Omega_{\rm Hor}=1/(2M)$, while the redshift factor diverges as expected from the approach to the horizon in the extremal limit. Nevertheless, the tidal corrections themselves become progressively suppressed, consistently with the general behavior discussed above. For retrograde orbits, both quantities remain finite as $\chi \to 1$.

\begin{figure}[ht]
    \centering
    \includegraphics[width=1.01\linewidth]{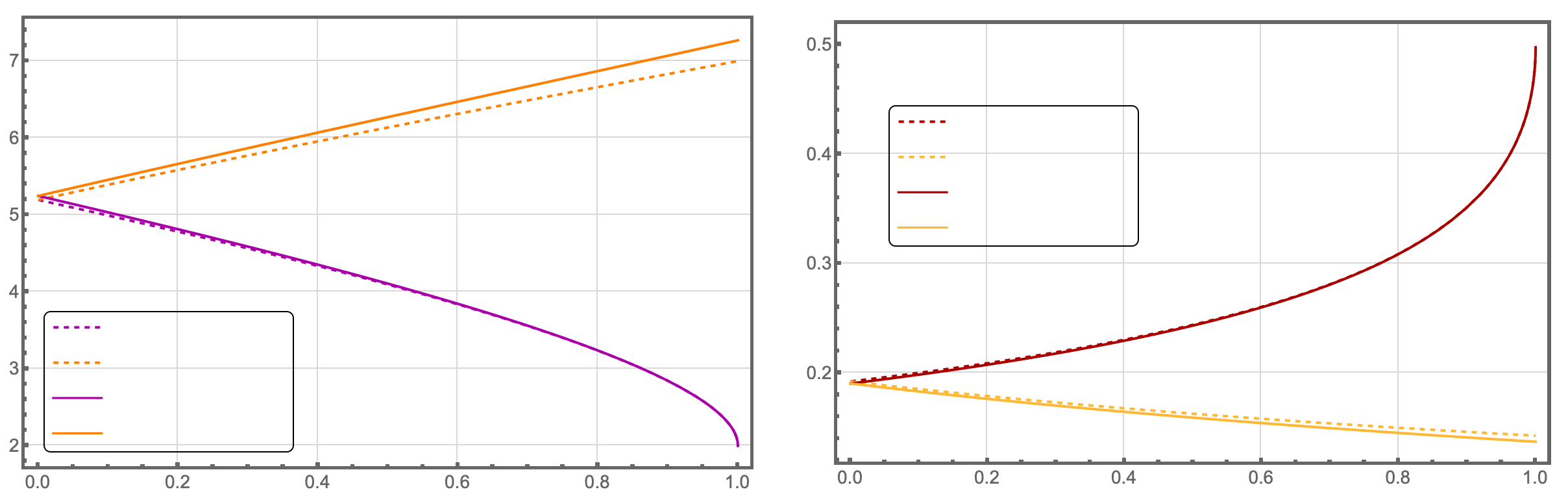}
    \put(-350,-5){\scriptsize $\chi$}
    \put(-115,-5){\scriptsize $\chi$}
    \put(-465,66){\rotatebox{90} {\scriptsize $b/M$}}
    \put(-230,66){\rotatebox{90} {\scriptsize $\Omega \, M$}}
     \put(-423,47){\tiny $\rm b_{0}$ ($\sigma$=+1)}
    \put(-423,37){\tiny $\rm b_{0}$ ($\sigma$=--1)}
    \put(-423,27){\tiny $\rm b_{0}$+$\eta$$\rm b_1$($\sigma$=+1)}
    \put(-423,17){\tiny $\rm b_{0}$+$\eta$$\rm b_1$($\sigma$=--1)}
     \put(-178,108){\tiny $\Omega_{0}$ ($\sigma$=+1)}
    \put(-179,98){\tiny |$\Omega_{0}$| ($\sigma$=--1)}
    \put(-178.5,87){\tiny $\Omega_{0}$+$\eta$$\Omega_1$($\sigma$=+1)}
    \put(-181,77){\tiny |$\Omega_{0}$+$\eta$$\Omega_1$|($\sigma$=--1)}
 \caption{LR impact parameter (left panel) and frequency (right panel), as functions of the dimensionless black-hole spin $\chi$. Dashed lines denote the unperturbed Kerr values, while solid lines denote the tidally corrected quantities $b_0^\sigma+\eta \, b_1^\sigma$ and $\Omega_0^\sigma+\eta \, \Omega_1^\sigma$. The impact parameter tends to the unperturbed value for prograde ($\sigma=+1$) orbits, while the deviation from the unperturbed value increases for retrograde ($\sigma=-1$) orbits. For retrograde motion, we display the absolute value of the frequency. We use $\eta=10^{-3}$ for visualization purposes.
}
\label{fig:b_and_Omega_LR_tidal_absolute}
\end{figure}

For the LR, the impact parameter behaves similarly to the radius, which is discussed extensively in Sec.~\ref{sec:ISCO_LR_shifts}.
The LR frequency behaves in a similar way to the ISCO case, as shown in Fig.~\ref{fig:b_and_Omega_LR_tidal_absolute}.
%

\clearpage
\bibliography{Bibliography}
\bibliographystyle{JHEP}


\end{document}